\input epsf
\documentstyle{amsppt}
\pagewidth{6.4truein}\hcorrection{0in}
\pageheight{9truein}\vcorrection{0.00in}

\TagsOnRight
\NoRunningHeads
\catcode`\@=11
\def\logo@{}
\footline={\ifnum\pageno>1 \hfil\folio\hfil\else\hfil\fi}
\topmatter
\title The emergence of the electrostatic field as a Feynman sum in random tilings with holes
\endtitle
\author Mihai Ciucu\endauthor
\thanks Research supported in part by NSF grant DMS 0500616.
\endthanks
\affil
Department of Mathematics, Indiana University\\
Bloomington, Indiana 47405
\endaffil
\abstract We consider random lozenge tilings on the triangular lattice with holes 
$Q_1,\dots,Q_n$ in some fixed position. For each unit triangle not in a hole, consider the
average orientation of the lozenge covering it. We show that the scaling limit of this discrete
field is the electrostatic field obtained when regarding each hole $Q_i$ as an 
electrical charge of magnitude equal to the difference between the number of unit triangles of
the two different orientations inside $Q_i$. This is then restated in terms of random
surfaces, yielding the result that the average over surfaces with prescribed
height at the union of the boundaries of the holes is, in the scaling limit, a sum of 
helicoids. 
\endabstract 
\endtopmatter
\document

\def\mysec#1{\bigskip\centerline{\bf #1}\message{ * }\nopagebreak\par\bigskip}

\def\myref#1{\item"{[{\bf #1}]}"} 
 
\def\pf{{\it Proof.\ }} 

\def\epf{\hfill{$\square$}\smallpagebreak}

\def\cite#1{\relaxnext@
  \def\nextiii@##1,##2\end@{[{\bf##1},\,##2]}%
  \in@,{#1}\ifin@\def\next{\nextiii@#1\end@}\else
  \def\next{[{\bf#1}]}\fi\next}
\def\proclaimheadfont@{\smc}

\def\pf{{\it Proof.\ }}

\define\Z{{\Bbb Z}}
\define\Q{{\Bbb Q}}
\define\R{{\Bbb R}}
\define\C{{\Bbb C}}
\define\z{\zeta}

\define\M{\operatorname{M}}
\define\Rep{\operatorname{Re}}

\define\ch{\operatorname{ch}}

\define\av{\operatorname{av}}
\define\twoline#1#2{\line{\hfill{\smc #1}\hfill{\smc #2}\hfill}}
\define\threeline#1#2#3{\line{\hfill{\smc #1}\hfill{\smc #2}\hfill{\smc #3}\hfill}}
\define\ltwoline#1#2{\line{{\smc #1}{\smc #2}}}

\def\mypic#1{\epsffile{figs/#1}}



\define\secA{4}

\define\ri{1}
\define\sc{2}
\define\ec{3}
\define\ov{4}
\define\CEP{5}
\define\CKP{6}
\define\CLP{7}
\define\ColdingNotices{8}
\define\QED{9}
\define\FS{10}
\define\Jackson{11}
\define\K{12}
\define\KOS{13}
\define\Sheffield{14}
\define\Thurst{15}

\vskip-0.05in
\mysec{Introduction}

The study of correlations of holes in random tilings was launched by Fisher and Stephenson 
\cite{\FS}, who considered in particular the monomer-monomer correlation and obtained exact data
suggesting rotational invariance in the scaling limit. Motivated by this, we studied 
correlations of a finite number of holes of various sizes and shapes in \cite{\ri}, \cite{\sc} and \cite{\ec},
and found that in the scaling limit they are given by a multiplicative version of the
superposition principle for energy in electrostatics. 
In this paper we consider the discrete field of average tile orientations in a random tiling with
holes, and provide the proof of the fact that the {\it electric field} 
comes about as the scaling limit of this discrete field. This result was announced 
in \cite{\ov}.


Consider the unit equilateral triangular lattice drawn in the plane so that some of the lattice 
lines are vertical. The union of any two unit triangles that share an edge is called a lozenge.
Inspired by Feynman's description of the reflection of light in \cite{\QED, Ch.\,2} and given the
connections between lozenge tilings with holes and electrostatics discussed in \cite{\sc} we construct
a discrete vector field as follows. 
Let $\Delta_1,\dotsc,\Delta_n$ be~holes in the lattice whose boundaries are
lattice triangles of even side-lengths. 
As one ranges over the set of lozenge tilings of the plane with these holes 
(a portion of such a tiling is illustrated in Figure 5.4(a)), any given left-pointing unit triangle $e$ in the 
complement of the holes is covered by a lozenge having one of three possible orientations: 
pointing in the polar direction 0, the polar direction $2\pi/3$, or the polar 
direction $4\pi/3$. 
Define ${\bold F}(e)$ to be the average of these orientations over all 
tilings (the precise definitions are given in the next section).


Now apply a homothety of modulus $1/R$ to our lattice with holes, and ``drag outward'' the 
images of the holes and the image of $e$ so that they shrink to points $z_1,\dotsc,z_n$ and
respectively $z_0$ in the complex plane, as $R\to\infty$.

Our main result can be phrased as follows.

\proclaim{Theorem} As $R\to\infty$ we have
$$
{\bold F}(e)\sim\frac{3}{4\pi R}\sum_{i=0}^n\ch(\Delta_i)
\frac{{\bold r}_{i0}}{|z_0-z_i|},
$$
where ${\bold r}_{i0}$ is the unit vector pointing from $z_i$ in the direction of $z_0$ and 
$\ch(Q)$ denotes the difference between the number of right- and left-pointing unit
triangles in the lattice region $Q$.
\endproclaim

The figure above shows on the left an instance of the field ${\bold F}$ in the case of two oppositely oriented 
holes of side two, and on the right the corresponding Coulomb field that it approaches in the scaling limit.

\topinsert
\twoline{\!\!\!\!\!\!\!\!\!\!\!\!\!\!\!\!\!\!\!\!\!\!\!\!\mypic{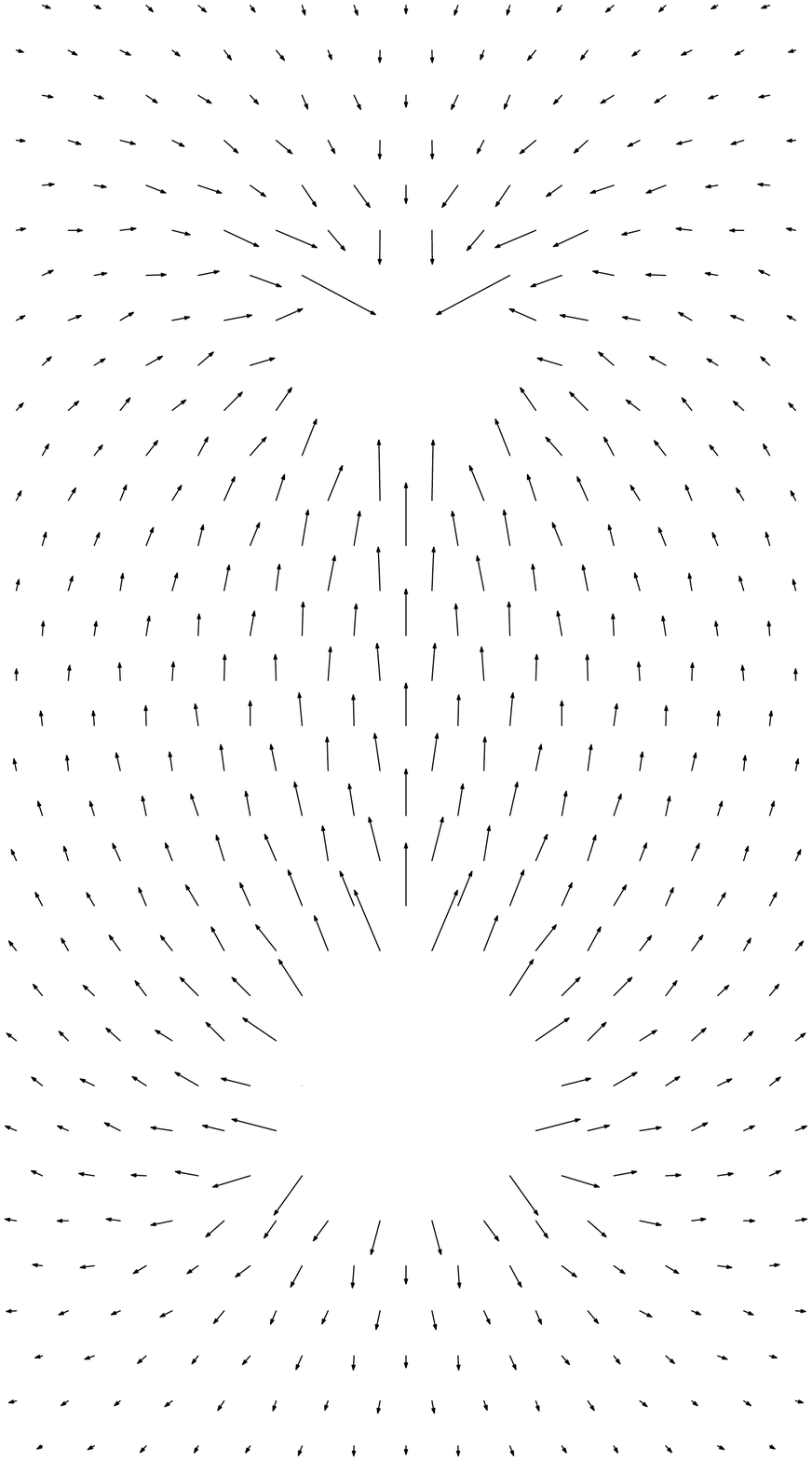}}
{\!\!\!\!\!\!\!\!\mypic{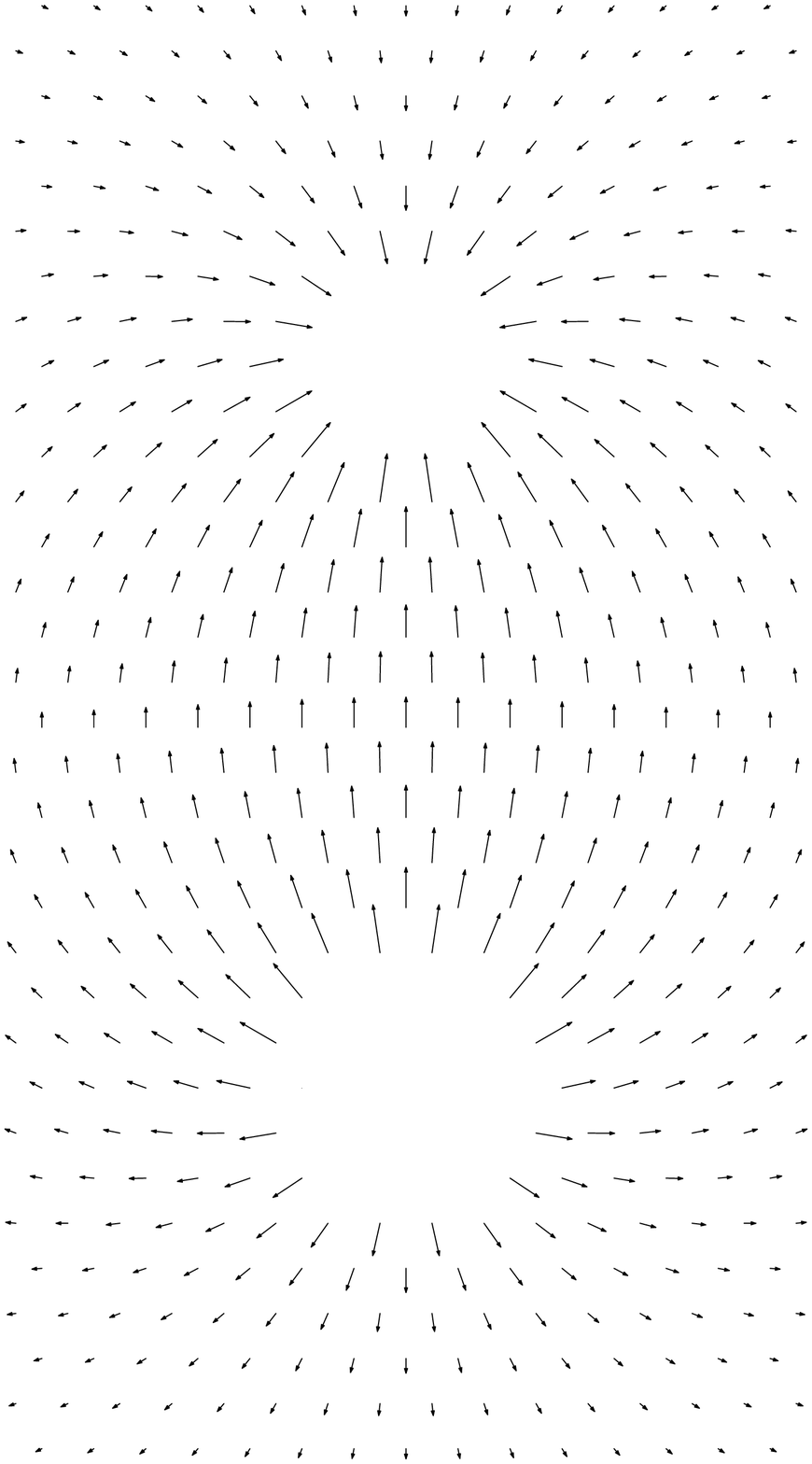}}
\endinsert

In Section 5 (see Theorem 5.1) we phrase this result in terms of random surfaces. Lozenge tilings of the plane 
with holes lift to certain discrete multi-sheeted surfaces. The above result implies that the average over
these surfaces (an instance of which is illustrated in Figure 5.5) converges in the scaling limit to a sum of 
helicoids (for lifting surfaces as in Figure 5.5, this limit surface is of the type depicted in Figure 5.7(b);
see \cite{\ColdingNotices} for an overview on a variety of settings in which helicoids arise).
The key to proving this is that we
obtain explicit formulas for the placement probabilities of lozenges in the scaling limit.

Phrased this way, our result is similar in spirit to the result of Cohn, Elkies and Propp \cite{\CEP} on the local 
statistics for random domino tilings of the Aztec diamond, where an explicit surface is found as the scaling limit
of the normalized height function. Another related instance of an explicit limiting surface,
concerning the placement probability of lozenges in a random tiling of a hexagon, was studied by Cohn, Larsen and 
Propp in \cite{\CLP}.
In \cite{\CKP}, Cohn, Kenyon and Propp show that the normalized average height function corresponding to domino 
tilings of regions that scale to an arbitrary simply connected region converges to the unique function that 
minimizes a certain surface tension integral. These authors also conjectured expressions for the local 
probability densities of domino configurations in large domino tilings. These conjectures were proved 
and generalized by Sheffield in \cite{\Sheffield}. 
Random surfaces corresponding to random dimer coverings of periodic bipartite planar lattices were studied
extensively by Kenyon, Okounkov and Sheffield in \cite{\KOS}. 

One notable difference between our result and the above mentioned ones is the presence of holes. This causes the
lifting surfaces to be multi-sheeted. The holes are also responsible for creating the field we are studying --- 
without them the field would be zero. 

It may also be worth
noting that for instance in \cite{\CLP} the boundary is the boundary of a lattice hexagon, and remains of fixed size
throughout the scaling process, while in our case the boundary consists of the union of finitely many boundaries
of triangular holes, each of them shrinking to a point in the scaling limit. Related to this is the fact that, unlike
the results from the literature mentioned above, we consider the average of un-normalized lifting surfaces --- the 
average of the normalized ones is identically zero.

\mysec{1. Definitions and statement of results}

Let $Q_j$ be a finite union of lattice triangular holes of side two with mutually disjoint 
interiors, for $j=1,\dotsc,n$ (even when $Q_j$ is not connected, we still view it as a single 
generalized hole, often referred to as a multihole). 
The joint correlation $\hat\omega(Q_1,\dotsc,Q_n)$ is defined in \cite{\ec} (and recalled in Section 5 of
the present paper)
by means of limits of tori; the definition is readily extended to allowing some of the $Q_j$'s to
be lozenges.
It follows from that definition that if $L$ is any possible
lozenge location, the probability that $L$ is occupied by a lozenge in a random lozenge
tiling with holes at $Q_1,\dotsc,Q_n$ is given by the ratio
$$
\frac{\hat\omega(L,Q_1,\dotsc,Q_n)}{\hat\omega(Q_1,\dotsc,Q_n)}
$$
(see Section 5 for the details).
It follows thus that the vector ${\bold F}(e)={\bold F}(e;Q_1,\dotsc,Q_n)$ described in the 
Introduction is given by
$$
{\bold F}(e)=\frac{\hat\omega(L_1,Q_1,\dotsc,Q_n)}{\hat\omega(Q_1,\dotsc,Q_n)}{\bold e}_1
+\frac{\hat\omega(L_2,Q_1,\dotsc,Q_n)}{\hat\omega(Q_1,\dotsc,Q_n)}{\bold e}_2
+\frac{\hat\omega(L_3,Q_1,\dotsc,Q_n)}{\hat\omega(Q_1,\dotsc,Q_n)}{\bold e}_3,\tag1.1
$$
where $L_1$, $L_2$ and $L_3$ are the lozenge locations containing the left-pointing unit triangle 
$e$ and pointing in the 
0, $2\pi/3$ and $4\pi/3$ polar directions, respectively, and the ${\bold e}_j$'s are unit 
vectors pointing in the directions of the long diagonals of the $L_j$'s.

Note that we can specify the location of any left- or 
right-pointing unit triangle (for short, we will call them left- and right-monomers) by indicating
the location of the midpoint of its vertical side. The midpoints of the vertical sides of the unit
triangles in our lattice can naturally be coordinatized by pairs of integers using a $60^\circ$ 
coordinate system with axes pointing in the polar directions $\pm\pi/3$.

Let $E_{a,b}$ be the east-pointing lattice triangle of side 2 whose central monomer
has coordinates $(a,b)$. Let $W_{a,b}$ be the similarly defined west-pointing lattice triangle. We
regard both of them as holes.

For any $q\in\Q$ and any strictly increasing list of integers ${\bold a}=(a_1,\dotsc,a_s)$ for
which $qa_i\in\Z$ and the $E_{a_i,qa_i}$'s (equivalently, the $W_{a_i,qa_i}$'s) are mutually 
disjoint, $i=1,\dotsc,s$, define the multiholes ${E}^q_{\bold a}$ and ${W}^q_{\bold a}$ by
$$
\align
{E}^q_{\bold a}&=E_{a_1,qa_1}\cup\dotsc\cup E_{a_s,qa_s}
\\
{W}^q_{\bold a}&=W_{a_1,qa_1}\cup\dotsc\cup W_{a_s,qa_s}. 
\endalign
$$
For a hole $Q$ in the lattice, let $Q(x,y)$ stand for its translation by the vector $(x,y)$ in our 
coordinate system. We say that an integer divides a rational number if it divides the numerator of
a lowest terms representation of it. 

We can now give the precise statement of our main result.

%

\proclaim{Theorem 1.1} Let $x_0^{(R)},\dotsc,x_m^{(R)}$, $y_0^{(R)},\dotsc,y_m^{(R)}$,
$z_0^{(R)},\dotsc,z_n^{(R)}$ and $w_0^{(R)},\dotsc,w_n^{(R)}$ be sequences of integers so that
$\lim_{R\to\infty}x_i^{(R)}/R=x_i$, $\lim_{R\to\infty}y_i^{(R)}/R=y_i$, 
$\lim_{R\to\infty}z_j^{(R)}/R=z_j$ and $\lim_{R\to\infty}w_j^{(R)}/R=w_j$ for $0\leq i\leq m$ and
$1\leq j\leq n$. Assume the $(x_i,y_i)$'s and $(z_j,w_j)$'s are all distinct. 

Then for any multiholes ${E}^q_{{\bold a}_1},\dotsc,{E}^q_{{\bold a}_m}$ and 
${W}^q_{{\bold b}_1},\dotsc,{W}^q_{{\bold b}_n}$ with $3|1-q$, the field 

$$
{\bold F}\left(x_0^{(R)},y_0^{(R)}\right)=
{\bold F}\left(x_0^{(R)},y_0^{(R)};E^q_{{\bold a}_1}(x_1^{(R)},y_1^{(R)}),
\dotsc,W^q_{{\bold b}_n}(z_n^{(R)},w_n^{(R)})\right)
$$ 
defined by $(1.1)$ has orthogonal projections on our coordinate axes with asymptotics 
$$
\align
&
F_x\left(x_0^{(R)},y_0^{(R)}\right)=\frac{3}{4\pi R}\left\{\sum_{i=1}^m s_i
\frac{2(x_0-x_i)+y_0-y_i}{(x_0-x_i)^2+(x_0-x_i)(y_0-y_i)+(y_0-y_i)^2}
\right.
\\
&
\left.
-\sum_{j=1}^n t_j
\frac{2(x_0-z_j)+y_0-w_j}{(x_0-z_j)^2+(x_0-z_j)(y_0-w_j)+(y_0-w_j)^2}\right\}
+o\left(\frac{1}{R}\right)\tag1.2
\endalign
$$
and 
$$ 
\align
&
F_y\left(x_0^{(R)},y_0^{(R)}\right)=\frac{3}{4\pi R}\left\{\sum_{i=1}^m s_i
\frac{x_0-x_i+2(y_0-y_i)}{(x_0-x_i)^2+(x_0-x_i)(y_0-y_i)+(y_0-y_i)^2}
\right.
\\
&
\left.
-\sum_{j=1}^n t_j
\frac{x_0-z_j+2(y_0-w_j)}{(x_0-z_j)^2+(x_0-z_j)(y_0-w_j)+(y_0-w_j)^2}\right\}
+o\left(\frac{1}{R}\right),\tag1.3
\endalign
$$
where $s_i$ and $t_j$ are the lengths of ${\bold a}_i$ and ${\bold b}_j$, respectively.

Furthermore, for any $\epsilon>0$ and any bounded set $B$ in the plane the implicit constants above are uniform
over all choices of the limits for which each distance among the points 
$(x_0,y_0),\dotsc,(z_n,w_n)\in B$ is at least~$\epsilon$.

\endproclaim 

Note that in our oblique coordinate system the Euclidean distance between the points
$(x,y)$ and $(x',y')$ is
$\sqrt{(x-x')^2+(x-x')(y-y')+(y-y')^2}$, and the orthogonal projections of $(x,y)$ on the coordinate
axes are $x+\frac12 y$ and $\frac12 x+y$. Note also that the multihole $E^1_{(0,2,4,\dotsc,2s-2)}$
consists of a contiguous horizontal string of right-pointing triangular holes of side two. Due to 
forced lozenges in its complement, $E^1_{(0,2,4,\dotsc,2s-2)}$ has precisely the same effect as 
the right-pointing triangular hole of side $2s$ that contains it; a similar statement holds for
$W$-multiholes.
Since $\ch(E^q_{{\bold a}_i})=2s_i$ and 
$\ch(W^q_{{\bold b}_j})=-2t_j$, one sees that the theorem stated in the Introduction follows as a 
special case of Theorem 1.1.

\mysec{2. Reducing the problem to exact determinant evaluations}

One crucial ingredient for proving the results of \cite{\ec} was an exact determinant formula 
for the joint correlation of an arbitrary collection of disjoint lattice-triangular holes of size two. 
The arguments presented there prove in fact a more general statement, which we will need in the current 
paper.

Our determinant formula involves the coupling function $P(x,y)$, $x,y\in\Z$ specified by
$$
P(x,y)=\frac{1}{2\pi i}\int_{e^{2\pi i/3}}^{e^{4\pi i/3}} t^{-y-1}(-1-t)^{-x-1}dt,\ \ \ x\leq-1\tag2.1
$$
and the symmetries $P(x,y)=P(y,x)=P(-x-y-1,x)$ (see Kenyon \cite{\K}),
and the coefficients $U_s$ of its asymptotic series
$$
P(-3r-1+a,-1+b)\sim\sum_{s=0}^\infty (3r)^{-s-1}U_s(a,b),\ \ \ r\to\infty,\ a,b\in\Z.\tag2.2
$$

Let $r(a,b)$ and $l(a,b)$ denote the right- and left-pointing monomers of coordinates $(a,b)$, 
respectively.

\proclaim{Proposition 2.1} Assume that $\{r(a_1,b_1),\dotsc,r(a_m,b_m),l(c_1,d_1),\dotsc,l(c_n,d_n)\}$
can be partitioned into subsets of size two so that the monomers in each subset share at least one 
vertex.
Then if $m\geq n$ we have
$$
\hat{\omega}(r(a_1,b_1),\dotsc,r(a_m,b_m),l(c_1,d_1),\dotsc,l(c_n,d_n))=
\left|\det\left[\matrix M_P\!\!\!\!\!\!\!\!&&M_U\endmatrix\right]\right|,
$$
where
$$
\spreadmatrixlines{2\jot}
\align
&
M_P=\left[
\matrix
P(a_1-c_1,b_1-d_1)&&\cdots&&P(a_1-c_n,b_1-d_n)\\
P(a_2-c_1,b_2-d_1)&&\cdots&&P(a_2-c_n,b_2-d_n)\\
\vdots&&\ &&\vdots\\
P(a_m-c_1,b_m-d_1)&&\cdots&&P(a_m-c_n,b_m-d_n)
\endmatrix\right]
\endalign
$$
and 
$$
\spreadmatrixlines{2\jot}
\align
M_U=
\left[
\matrix
U_0(a_1,b_1+1)\!\!\!\!\!\!\!\!&&U_0(a_1+1,b_1)\!\!\!\!&&\cdots\!\!\!\!&&
U_{\frac{m-n}{2}-1}(a_1,b_1+1)\!\!\!\!\!\!\!\!&&U_{\frac{m-n}{2}-1}(a_1+1,b_1)\\
U_0(a_2,b_2+1)\!\!\!\!\!\!\!\!&&U_0(a_2+1,b_2)\!\!\!\!&&\cdots\!\!\!\!&&
U_{\frac{m-n}{2}-1}(a_2,b_2+1)\!\!\!\!\!\!\!\!&&U_{\frac{m-n}{2}-1}(a_2+1,b_2)\\
\vdots&&\vdots&&\ &&\vdots&&\vdots\\
U_0(a_m,b_m+1)\!\!\!\!\!\!\!\!&&U_0(a_m+1,b_m)\!\!\!\!&&\cdots\!\!\!\!&&
U_{\frac{m-n}{2}-1}(a_m,b_m+1)\!\!\!\!\!\!\!\!&&U_{\frac{m-n}{2}-1}(a_m+1,b_m)
\endmatrix\right]\!\!.
\endalign
$$

\endproclaim

\pf Since the set of monomers can be partitioned into pairs, $m+n$ is even; hence so is $m-n$. 
Thus we can prove the statement by induction on $m-n$ as follows. 
To relate a given configuration to one for which 
$m-n$ is 2 units smaller, bring in and extra hole, $W(3r,0)$, and let $r\to\infty$. The details go 
through just as in the proof of \cite{\ec,Proposition~3.2}.  \epf

In order to obtain the asymptotics of the field ${\bold F}(e)$ we need to understand
the three coefficients in (1.1) when the holes are $E_{{\bold a}_1},\dotsc,E_{{\bold a}_m},
W_{{\bold b}_1},\dotsc,W_{{\bold b}_n}$.
The asymptotics of the denominators in the special case $x_1^{(R)}=Rx_1,\dotsc,w_n^{(R)}=Rw_n$, 
$x_1,\dotsc,w_n\in3\Z$,
is worked out in \cite{\ec,Theorem~8.1}. Proposition 2.1 supplies exact determinant expressions for the numerators.
Each is the determinant of a block matrix whose blocks consist in turn of (mostly) $2\times2$ blocks.
To write them down it will be helpful to define the following five families of matrices.

Given two lists ${\bold a}=(a_1,\dotsc,a_s)$ and ${\bold b}=(b_1,\dotsc,b_t)$, let 
$$
\align
A_{x,y,z,w}({\bold a},{\bold b})=
\left[\matrix
{\scriptstyle \cdots}&&
{\scriptstyle P(x-z+a_1-b_j-1,y-w+q(a_1-b_j)-1)}\!\!\!\!&&{\scriptstyle P(x-z+a_1-b_j-2,y-w+q(a_1-b_j))}
&&{\scriptstyle \cdots}
\\
{\scriptstyle \cdots}&&
{\scriptstyle P(x-z+a_1-b_j,y-w+q(a_1-b_j)-2)}\!\!\!\!&&{\scriptstyle P(x-z+a_1-b_j-1,y-w+q(a_1-b_j)-1)}
&&{\scriptstyle \cdots}
\\
\\
{\scriptstyle \cdots}&&
{\scriptstyle P(x-z+a_2-b_j-1,y-w+q(a_2-b_j)-1)}\!\!\!\!&&{\scriptstyle P(x-z+a_2-b_j-2,y-w+q(a_2-b_j))}
&&{\scriptstyle \cdots}
\\
{\scriptstyle \cdots}&&
{\scriptstyle P(x-z+a_2-b_j,y-w+q(a_2-b_j)-2)}\!\!\!\!&&{\scriptstyle P(x-z+a_2-b_j-1,y-w+q(a_2-b_j)-1)}
&&{\scriptstyle \cdots}
\\
\\
{\scriptstyle \vdots}&&{\scriptstyle \vdots}\!\!\!\!&&{\scriptstyle \vdots}&&{\scriptstyle \vdots}
\\
\\
{\scriptstyle \cdots}&&
{\scriptstyle P(x-z+a_s-b_j-1,y-w+q(a_s-b_j)-1)}\!\!\!\!&&{\scriptstyle P(x-z+a_s-b_j-2,y-w+q(a_s-b_j))}
&&{\scriptstyle \cdots}
\\
{\scriptstyle \cdots}&&
{\scriptstyle P(x-z+a_s-b_j,y-w+q(a_s-b_j)-2)}\!\!\!\!&&{\scriptstyle P(x-z+a_s-b_j-1,y-w+q(a_s-b_j)-1)}
&&{\scriptstyle \cdots}
\endmatrix\right]
\\
\tag2.3
\endalign
$$
(the display shows the $j$th ``column'' of $2\times2$ blocks of $A_{x,y,z,w}({\bold a},{\bold b})$;
there is one for each $j=1,\dotsc,t$).

The second family of matrices is
$$
\align
B_{x,y}({\bold a},k)=
\left[\matrix
{\scriptstyle U_0(x+a_1,y+qa_1)}\!\!\!\!\!\!\!\!\!\!&&{\scriptstyle U_0(x+a_1-1,y+qa_1+1)}\!\!\!\!\!\!\!\!\!\!&&
{\scriptstyle \cdots}\!\!\!\!\!\!\!\!\!\!&&
{\scriptstyle U_{k}(x+a_1,y+qa_1)}\!\!\!\!\!\!\!\!\!\!&&{\scriptstyle U_{k}(x+a_1-1,y+qa_1+1)}
\\
{\scriptstyle U_0(x+a_1+1,y+qa_1-1)}\!\!\!\!\!\!\!\!\!\!&&{\scriptstyle U_0(x+a_1,y+qa_1)}\!\!\!\!\!\!\!\!\!\!&&
{\scriptstyle \cdots}\!\!\!\!\!\!\!\!\!\!&&
{\scriptstyle U_{k}(x+a_1+1,y+qa_1-1)}\!\!\!\!\!\!\!\!\!\!&&{\scriptstyle U_{k}(x+a_1,y+qa_1)}
\\
\\
{\scriptstyle U_0(x+a_2,y+qa_2)}\!\!\!\!\!\!\!\!\!\!&&{\scriptstyle U_0(x+a_2-1,y+qa_2+1)}\!\!\!\!\!\!\!\!\!\!&&
{\scriptstyle \cdots}\!\!\!\!\!\!\!\!\!\!&&
{\scriptstyle U_{k}(x+a_2,y+qa_2)}\!\!\!\!\!\!\!\!\!\!&&{\scriptstyle U_{k}(x+a_2-1,y+qa_2+1)}
\\
{\scriptstyle U_0(x+a_2+1,y+qa_2-1)}\!\!\!\!\!\!\!\!\!\!&&{\scriptstyle U_0(x+a_2,y+qa_2)}\!\!\!\!\!\!\!\!\!\!&&
{\scriptstyle \cdots}\!\!\!\!\!\!\!\!\!\!&&
{\scriptstyle U_{k}(x+a_2+1,y+qa_2-1)}\!\!\!\!\!\!\!\!\!\!&&{\scriptstyle U_{k}(x+a_2,y+qa_2)}
\\
\\
{\scriptstyle \vdots}\!\!\!\!\!\!\!\!\!\!&&{\scriptstyle \vdots}\!\!\!\!\!\!\!\!\!\!&&
{\scriptstyle \ }\!\!\!\!\!\!\!\!\!\!&&{\scriptstyle \vdots}\!\!\!\!\!\!\!\!\!\!&&
{\scriptstyle \vdots}
\\
\\
{\scriptstyle U_0(x+a_s,y+qa_s)}\!\!\!\!\!\!\!\!\!\!&&{\scriptstyle U_0(x+a_s-1,y+qa_s+1)}\!\!\!\!\!\!\!\!\!\!&&
{\scriptstyle \cdots}\!\!\!\!\!\!\!\!\!\!&&
{\scriptstyle U_{k}(x+a_s,y+qa_s)}\!\!\!\!\!\!\!\!\!\!&&{\scriptstyle U_{k}(x+a_s-1,y+qa_s+1)}
\\
{\scriptstyle U_0(x+a_s+1,y+qa_s-1)}\!\!\!\!\!\!\!\!\!\!&&{\scriptstyle U_0(x+a_s,y+qa_s)}\!\!\!\!\!\!\!\!\!\!&&
{\scriptstyle \cdots}\!\!\!\!\!\!\!\!\!\!&&
{\scriptstyle U_{k}(x+a_s+1,y+qa_s-1)}\!\!\!\!\!\!\!\!\!\!&&{\scriptstyle U_{k}(x+a_s,y+qa_s)}
\endmatrix\right].
\\
\tag2.4
\endalign
$$
The remaining families consist of one-row or one-column matrices. The third is
$$
C_{x,y,z,w}({\bold b})=
\left[\matrix
{\scriptstyle \cdots}&&
{\scriptstyle P(x-z-b_j,y-w-qb_j-1)}\!\!\!\!&&{\scriptstyle P(x-z-b_j-1,y-w-qb_j)}
&&{\scriptstyle \cdots}
\endmatrix\right].
\tag2.5
$$
The fourth family is
$$
D_{x,y,z,w}({\bold a})=
\left[\matrix
P(x-z+a_1-1,y-w+qa_1)\\
P(x-z+a_1,y-w+qa_1-1)\\
\\
P(x-z+a_2-1,y-w+qa_2)\\
P(x-z+a_2,y-w+qa_2-1)\\
\\
\vdots\\
\\
P(x-z+a_s-1,y-w+qa_s)\\
P(x-z+a_s,y-w+qa_s-1)
\endmatrix\right],
\tag2.6
$$
and the fifth
$$
G_{x,y}(k)=
\left[\matrix
{ U_0(x+1,y)}\!\!\!\!\!\!\!\!\!\!&&{ U_0(x,y+1)}\!\!\!\!\!\!\!\!\!\!&&
{ \cdots}\!\!\!\!\!\!\!\!\!\!&&
{ U_{k}(x+1,y)}\!\!\!\!\!\!\!\!\!\!&&{ U_{k}(x,y+1)}
\endmatrix\right].
\tag2.7
$$

Let $L_1(x,y)$, $L_2(x,y)$, and $L_3(x,y)$ be the the lozenge locations pointing in the
polar directions 0, $2\pi/3$, $4\pi/3$, respectively, and containing the left-monomer $(x,y)$. 
Let $s_i$ and $t_j$ be the lengths of the lists ${\bold a}_i$ and~${\bold b}_j$, respectively, for $i=1,\dotsc,m$,
$j=1,\dotsc,n$.
Then Proposition 2.1 implies\footnote{ As in \cite{\ec,Lemma 5.1}, we use the fact that $E(x,y)$ can
be replaced by the union of the two right-monomers $r(x-1,y)$ and $r(x,y-1)$ , and $W(x,y)$
by $l(x+1,y)\cup l(x,y+1)$.}
$$
\hat\omega\left(L_1\left(x_0^{(R)},y_0^{(R)}\right),
E_{{\bold a}_1}^q\left(x_1^{(R)},y_1^{(R)}\right),\dotsc,
W_{{\bold b}_n}^q\left(z_n^{(R)},w_n^{(R)}\right)\right)=\left|\det \bar{M}_1\right|,\tag2.8
$$
where
$$
\spreadlines{4\jot}
\align
&
\bar{M}_1=
\\
&\ 
\left[\!\matrix
P(0,0)\!\!\!\!\!\!\!\!&&
C_{x_0^{(R)},y_0^{(R)},z_1^{(R)},w_1^{(R)}}({\bold b}_1)\!\!\!\!\!\!\!\!&&\cdots\!\!\!\!\!\!\!\!&&
C_{x_0^{(R)},y_0^{(R)},z_n^{(R)},w_n^{(R)}}({\bold b}_n)\!\!\!\!\!\!\!\!&&
G_{x_0^{(R)},y_0^{(R)}}(\nu)
\\
\\
\\
D_{x_1^{(R)},y_1^{(R)},x_0^{(R)},y_0^{(R)}}({\bold a}_1)\!\!\!\!\!\!\!\!&&
A_{x_1^{(R)},y_1^{(R)},z_1^{(R)},w_1^{(R)}}({\bold a}_1,{\bold b}_1)\!\!\!\!\!\!\!\!&&\cdots\!\!\!\!\!\!\!\!&&
A_{x_1^{(R)},y_1^{(R)},z_n^{(R)},w_n^{(R)}}({\bold a}_1,{\bold b}_n)\!\!\!\!\!\!\!\!&&
B_{x_1^{(R)},y_1^{(R)}}({\bold a}_1,\nu)
\\
\\
\\
\vdots\!\!\!\!\!\!\!\!&&\vdots\!\!\!\!\!\!\!\!&&\ \!\!\!\!\!\!\!\!&&\vdots\!\!\!\!\!\!\!\!&&\vdots
\\
\\
\\
D_{x_m^{(R)},y_m^{(R)},x_0^{(R)},y_0^{(R)}}({\bold a}_m)\!\!\!\!\!\!\!\!&&
A_{x_m^{(R)},y_m^{(R)},z_1^{(R)},w_1^{(R)}}({\bold a}_m,{\bold b}_1)\!\!\!\!\!\!\!\!&&\cdots\!\!\!\!\!\!\!\!&&
A_{x_m^{(R)},y_m^{(R)},z_n^{(R)},w_n^{(R)}}({\bold a}_m,{\bold b}_n)\!\!\!\!\!\!\!\!&&
B_{x_m^{(R)},y_m^{(R)}}({\bold a}_m,\nu)
\endmatrix\!\right]\!\!,
\\
\tag2.9
\endalign
$$
and $\nu=\sum_{i=1}^m s_i - \sum_{j=1}^n t_j-1$. We set $S=s_1+\cdots+s_m$ and $T=t_1+\cdots+t_n$.

Another application of Proposition 2.1 gives
$$
\hat\omega\left(
E_{{\bold a}_1}^q\left(x_1^{(R)},y_1^{(R)}\right),\dotsc,
W_{{\bold b}_n}^q\left(z_n^{(R)},w_n^{(R)}\right)\right)=|\det M|,\tag2.10
$$
where\footnote{\, We write $A_I^J$ for the submatrix of $A$ at the intersection of the rows
with indices in $I$ with the columns with indices in $J$; $[n]$ stands for $\{1,\dotsc,n\}$.}
$$
M=\left(\bar{M}_1\right)_{[2S+1]\setminus\{1\}}^{[2S+1]\setminus\{1\}}.\tag2.11
$$

Denote 
$$
p_1:=\frac{
\hat\omega\left(L_1\left(x_0^{(R)},y_0^{(R)}\right),
E_{{\bold a}_1}^q\left(x_1^{(R)},y_1^{(R)}\right),\dotsc,
W_{{\bold b}_n}^q\left(z_n^{(R)},w_n^{(R)}\right)\right)
}
{
\hat\omega\left(
E_{{\bold a}_1}^q\left(x_1^{(R)},y_1^{(R)}\right),\dotsc,
W_{{\bold b}_n}^q\left(z_n^{(R)},w_n^{(R)}\right)\right)
}.\tag2.12
$$

Throughout this paper $\z=e^{2\pi i/3}$.

\proclaim{Proposition 2.2} If $x_0^{(R)}/R\to x_0,\dotsc,w_n^{(R)}/R\to w_n$ and
$x_0^{(R)}=\alpha_0\,(\text{\rm mod}\ 3),\dotsc,w_n^{(R)}=\delta_n\,(\text{\rm mod}\,\, 3)$, we have
$$
p_1=\frac{1}{3}
+\frac{1}{2\pi i R}\ \frac{\det M_1''}{\det M''}
+o\left(\frac{1}{R}\right),\tag2.13
$$
where 
$$
\spreadlines{4\jot}
\align
&
M_1''=
\\
&
\left[\!\matrix
0\!\!&&
\dot{C}_{x_0,y_0,z_1,w_1}^{\alpha_0,\beta_0,\gamma_1,\delta_1}(t_1)
\!\!&&
\cdots\!\!&&
\dot{C}_{x_0,y_0,z_n,w_n}^{\alpha_0,\beta_0,\gamma_n,\delta_n}(t_n)
\!\!&&
\dot{G}_{x_0,y_0}^{\alpha_0,\beta_0}(S-T-1)
\\
\\
\\
\dot{D}_{x_1,y_1,x_0,y_0}^{\alpha_1,\beta_1,\alpha_0,\beta_0}(s_1)\!\!&&
\dot{A}_{x_1,y_1,z_1,w_1}^{\alpha_1,\beta_1,\gamma_1,\delta_1}(s_1,t_1)\!\!&&\cdots\!\!&&
\dot{A}_{x_1,y_1,z_n,w_n}^{\alpha_1,\beta_1,\gamma_n,\delta_n}(s_1,t_n)\!\!&&
\dot{B}_{x_1,y_1}^{\alpha_1,\beta_1}(s_1,S-T-1)
\\
\\
\\
\vdots\!\!&&\vdots\!\!&&\ \!\!&&\vdots\!\!&&\vdots
\\
\\
\\
\dot{D}_{x_m,y_m,x_0,y_0}^{\alpha_m,\beta_m,\alpha_0,\beta_0}(s_m)\!\!&&
\dot{A}_{x_m,y_m,z_1,w_1}^{\alpha_m,\beta_m,\gamma_1,\delta_1}(s_m,t_1)\!\!&&
\cdots\!\!&&
\dot{A}_{x_m,y_m,z_n,w_n}^{\alpha_m,\beta_m,\gamma_n,\delta_n}(s_m,t_n)\!\!&&
\dot{B}_{x_m,y_m}^{\alpha_m,\beta_m}(s_m,S-T-1)
\endmatrix\!\right]\!\!
\\
\tag2.14
\endalign
$$
and
$$
M''=\left(M_1''\right)_{[2S+1]\setminus\{1\}}^{[2S+1]\setminus\{1\}},\tag2.15
$$
with\footnote{\,
Here and throughout the rest of the paper $\langle f(\z)\rangle$ stands for $f(\z)-f(\z^{-1})$.
}
$$
\spreadmatrixlines{1\jot}
\dot{A}_{x,y,z,w}^{\alpha,\beta,\gamma,\delta}(s,t)=
\left[\matrix
{ \cdots}&&
{ \left\langle\frac{ \z^{-1+(\alpha-\beta)-(\gamma-\delta)}{j-1\choose j-1}(1-q\z)^{j-1} }
{ [z-x-(w-y)\z]^{j} }\right\rangle }
\!\!\!\!\!\!\!\!&&
{ \left\langle\frac{\z^{-3+(\alpha-\beta)-(\gamma-\delta)}{j-1\choose j-1}(1-q\z)^{j-1}}
{[z-x-(w-y)\z]^{j}}\right\rangle}
&&{ \cdots}
\\
{ \cdots}&&
{ \left\langle\frac{\z^{1+(\alpha-\beta)-(\gamma-\delta)}{j-1\choose j-1}(1-q\z)^{j-1}}
{[z-x-(w-y)\z]^{j}}\right\rangle}
\!\!\!\!\!\!\!\!&&
{ \left\langle\frac{\z^{-1+(\alpha-\beta)-(\gamma-\delta)}{j-1\choose j-1}(1-q\z)^{j-1}}
{[z-x-(w-y)\z]^{j}}\right\rangle}
&&{ \cdots}
\\
\\
\\
{ \cdots}&&
{ \left\langle\frac{\z^{-1+(\alpha-\beta)-(\gamma-\delta)}{j\choose j-1}(1-q\z)^{j} }
{[z-x-(w-y)\z]^{j+1}}\right\rangle}
\!\!\!\!\!\!\!\!&&
{ \left\langle\frac{\z^{-3+(\alpha-\beta)-(\gamma-\delta)}{j\choose j-1}(1-q\z)^{j}}
{[z-x-(w-y)\z]^{j+1}}\right\rangle}
&&{ \cdots}
\\
{ \cdots}&&
{ \left\langle\frac{\z^{1+(\alpha-\beta)-(\gamma-\delta)}{j\choose j-1}(1-q\z)^{j}}
{[z-x-(w-y)\z]^{j+1}}\right\rangle}
\!\!\!\!\!\!\!\!&&
{ \left\langle\frac{\z^{-1+(\alpha-\beta)-(\gamma-\delta)}{j\choose j-1}(1-q\z)^{j}}
{[z-x-(w-y)\z]^{j+1}}\right\rangle}
&&{ \cdots}
\\
\\
{ \ }&&{ \vdots}
\!\!\!\!\!\!\!\!&&{ \vdots}&&{ \ }
\\
\\
{ \cdots}&&
{ \left\langle\frac{\z^{-1+(\alpha-\beta)-(\gamma-\delta)}{s+j-2\choose j-1}(1-q\z)^{s+j-2}}
{[z-x-(w-y)\z]^{s+j-1}}\right\rangle}
\!\!\!\!\!\!\!\!&&
{ \left\langle\frac{\z^{-3+(\alpha-\beta)-(\gamma-\delta)}{s+j-2\choose j-1}(1-q\z)^{s+j-2}}
{[z-x-(w-y)\z]^{s+j-1}}\right\rangle}
&&{ \cdots}
\\
{ \cdots}&&
{ \left\langle\frac{\z^{1+(\alpha-\beta)-(\gamma-\delta)}{s+j-2\choose j-1}(1-q\z)^{s+j-2}}
{[z-x-(w-y)\z]^{s+j-1}}\right\rangle}
\!\!\!\!\!\!\!\!&&
{ \left\langle\frac{\z^{-1+(\alpha-\beta)-(\gamma-\delta)}{s+j-2\choose j-1}(1-q\z)^{s+j-2}}
{[z-x-(w-y)\z]^{s+j-1}}\right\rangle}
&&{ \cdots}
\endmatrix\right]
\tag2.16
$$
$($there are $t$ pairs of such columns, one for each $j=1,\dotsc,t$$)$,
$$
\spreadlines{3\jot}
\spreadmatrixlines{1\jot}
\align
&
\dot{B}_{x,y}^{\alpha,\beta}(s,k)=
\\
&
\left[\!\!\matrix
\left\langle\frac{ \z^{-1+\alpha-\beta}{0\choose 0}(1-q\z)^0}{(x-y\z)^0}\right\rangle\!\!\!\!\!\!\!\!\!\!\!&&
\left\langle\frac{\z^{-3+\alpha-\beta}{0\choose 0}(1-q\z)^0}{(x-y\z)^0}\right\rangle\!\!\!\!\!\!\!\!\!\!\!&&
\cdots\!\!\!\!\!\!\!\!\!\!\!&&
\left\langle\frac{\z^{-1+\alpha-\beta}{k\choose 0}(1-q\z)^0}{(x-y\z)^{-k}}\right\rangle\!\!\!\!\!\!\!\!\!\!\!&&
\left\langle\frac{\z^{-3+\alpha-\beta}{k\choose 0}(1-q\z)^0}{(x-y\z)^{-k}}\right\rangle
\\
\left\langle\frac{ \z^{1+\alpha-\beta}{0\choose 0}(1-q\z)^0}{(x-y\z)^0}\right\rangle\!\!\!\!\!\!\!\!\!\!\!&&
\left\langle\frac{\z^{-1+\alpha-\beta}{0\choose 0}(1-q\z)^0}{(x-y\z)^0}\right\rangle\!\!\!\!\!\!\!\!\!\!\!&&
\cdots\!\!\!\!\!\!\!\!\!\!\!&&
\left\langle\frac{\z^{1+\alpha-\beta}{k\choose 0}(1-q\z)^0}{(x-y\z)^{-k}}\right\rangle\!\!\!\!\!\!\!\!\!\!\!&&
\left\langle\frac{\z^{-1+\alpha-\beta}{k\choose 0}(1-q\z)^0}{(x-y\z)^{-k}}\right\rangle
\\
\\
\vdots\!\!\!\!\!\!\!\!\!\!\!&&\!\!\!\!\!\!\!\!\!\!\!&&
\ \!\!\!\!\!\!\!\!\!\!\!&&
\vdots\!\!\!\!\!\!\!\!\!\!\!&&\vdots
\\
\\
\left\langle\frac{\z^{-1+\alpha-\beta}{0\choose s-1}(1-q\z)^{s-1}}{(x-y\z)^{s-1}}
\right\rangle
\!\!\!\!\!\!\!\!\!\!\!&&
\left\langle\frac{\z^{-3+\alpha-\beta}{0\choose s-1}(1-q\z)^{s-1}}{(x-y\z)^{s-1}}
\right\rangle
\!\!\!\!\!\!\!\!\!\!\!&&
\cdots\!\!\!\!\!\!\!\!\!\!\!&&
\left\langle\frac{\z^{-1+\alpha-\beta}{k\choose s-1}(1-q\z)^{s-1}}{(x-y\z)^{s-1-k}}
\right\rangle
\!\!\!\!\!\!\!\!\!\!\!&&
\left\langle\frac{\z^{-3+\alpha-\beta}{k\choose s-1}(1-q\z)^{s-1}}{(x-y\z)^{s-1-k}}
\right\rangle
\\
\left\langle\frac{\z^{1+\alpha-\beta}{0\choose s-1}(1-q\z)^{s-1}}{(x-y\z)^{s-1}}
\right\rangle
\!\!\!\!\!\!\!\!\!\!\!&&
\left\langle\frac{\z^{-1+\alpha-\beta}{0\choose s-1}(1-q\z)^{s-1}}{(x-y\z)^{s-1}}
\right\rangle
\!\!\!\!\!\!\!\!\!\!\!&&
\cdots\!\!\!\!\!\!\!\!\!\!\!&&
\left\langle\frac{\z^{1+\alpha-\beta}{k\choose s-1}(1-q\z)^{s-1}}{(x-y\z)^{s-1-k}}
\right\rangle
\!\!\!\!\!\!\!\!\!\!\!&&
\left\langle\frac{\z^{-1+\alpha-\beta}{k\choose s-1}(1-q\z)^{s-1}}{(x-y\z)^{s-1-k}}\right\rangle
\endmatrix\!\!\right]\!\!,
\\
\tag2.17
\endalign
$$
$$
\dot{C}_{x,y,z,w}^{\alpha,\beta,\gamma,\delta}(t)=
\left[\matrix
{ \cdots}&&
{ \left\langle\frac{ \z^{0+(\alpha-\beta)-(\gamma-\delta)}(1-q\z)^{j-1} }
{ [z-x-(w-y)\z]^{j} }\right\rangle }
\!\!\!\!\!\!\!\!&&
{ \left\langle\frac{\z^{-2+(\alpha-\beta)-(\gamma-\delta)}(1-q\z)^{j-1}}
{[z-x-(w-y)\z]^{j}}\right\rangle}
&&{ \cdots}
\endmatrix\right],\tag2.18
$$
$$
\spreadmatrixlines{1\jot}
\dot{D}_{x,y,z,w}^{\alpha,\beta,\gamma,\delta}(s)=
\left[\matrix
\left\langle\frac{\z^{-2+(\alpha-\beta)-(\gamma-\delta)}(1-q\z)^0}{[z-x-(w-y)\z]}\right\rangle\\
\left\langle\frac{\z^{0+(\alpha-\beta)-(\gamma-\delta)}(1-q\z)^0}{[z-x-(w-y)\z]}\right\rangle\\
\\
\left\langle\frac{\z^{-2+(\alpha-\beta)-(\gamma-\delta)}(1-q\z)^1}{[z-x-(w-y)\z]^2}\right\rangle\\
\left\langle\frac{\z^{0+(\alpha-\beta)-(\gamma-\delta)}(1-q\z)^1}{[z-x-(w-y)\z]^2}\right\rangle\\
\\
\vdots\\
\\
\left\langle\frac{\z^{-2+(\alpha-\beta)-(\gamma-\delta)}(1-q\z)^{s-1}}{[z-x-(w-y)\z]^s}\right\rangle\\
\left\langle\frac{\z^{0+(\alpha-\beta)-(\gamma-\delta)}(1-q\z)^{s-1}}{[z-x-(w-y)\z]^s}\right\rangle
\endmatrix\right],
\tag2.19
$$
and
$$
\dot{G}_{x,y}^{\alpha,\beta}(k)=
\left[\matrix
\left\langle\frac{\z^{0+\alpha-\beta}}{(x-y\z)^0}\right\rangle
\!\!\!\!\!\!\!\!\!\!&&
\left\langle\frac{\z^{-2+\alpha-\beta}}{(x-y\z)^0}\right\rangle
\!\!\!\!\!\!\!\!\!\!&&
\left\langle\frac{\z^{0+\alpha-\beta}}{(x-y\z)^{-1}}\right\rangle
\!\!\!\!\!\!\!\!\!\!&&
\left\langle\frac{\z^{-2+\alpha-\beta}}{(x-y\z)^{-1}}\right\rangle
\!\!\!\!\!\!\!\!\!\!&&
{ \cdots}\!\!\!\!\!\!\!\!\!\!&&
\left\langle\frac{\z^{0+\alpha-\beta}}{(x-y\z)^{-k}}\right\rangle
\!\!\!\!\!\!\!\!\!\!&&
\left\langle\frac{\z^{-2+\alpha-\beta}}{(x-y\z)^{-k}}\right\rangle
\endmatrix\right].
\tag2.20
$$
Furthermore, for any $\epsilon>0$ and any bounded set $B$, the implicit constant in $(2.13)$ is uniform
for all choices of $x_0,\dotsc,w_n$ for which $(x_0,y_0),\dotsc,(z_n,w_n)\in B$ have all
mutual distances at least $\epsilon$.

\endproclaim

Note that (2.13)--(2.20) (plus 2 analogs) imply already, before evaluating the determinants, 
that in the scaling limit ${\bold F}(e)$ does not depend on the spacings between the side-two 
constituent holes of the multiholes $E^q_{{\bold a}_i}$ and $W^q_{{\bold b}_j}$ (although interestingly 
the independence of $q$ is not yet apparent). 

\medskip
\pf By (2.12), (2.8) and (2.10) we have
$$
p_1=\left|\frac{\det \bar{M}_1}{\det M}\right|.\tag2.21
$$
It readily follows from the integral expression (2.1) for $P$ that $P(0,0)=1/3$. Regarding the 
first column of $\bar{M}_1$ as the sum of two vectors
one of which is $(1/3,0,\dotsc,0)^{T}$ and using the linearity of the determinant yields
$$
p_1=\left|\frac{1}{3}+\frac{\det M_1}{\det M}\right|,\tag2.22
$$
where 
$$
M_1=\bar{M}_1|_{(1,1)\text{\rm-entry set to $0$}}.\tag2.23
$$

In the special case $x_1^{(R)}=Rx_1,\dotsc,w_n^{(R)}=Rw_n$ and $x_0=\cdots=w_n=0\,(\text{\rm mod}\,\, 3)$,
Lemma 5.3 of \cite{ec} shows how to write the main term in the asymptotics of $\det M$ as a determinant 
independent of $R$. This can be accomplished in the current case of
general $x_1^{(R)},\dotsc,w_n^{(R)}$ by essentially the same arguments. 

Namely, consider the following operation on a square matrix $X$ in which rows 
$i_1,\dotsc,i_k$ are of the form
$f(c_1),\dotsc,f(c_k)$, respectively, for some vector function $f$: Transform
rows $i_1,\dotsc,i_k$ of $X$ as
$$
\left[\matrix
f(c_1)\\
f(c_2)\\
.\\
.\\
.\\
f(c_k)
\endmatrix\right]
\mapsto
\left[\matrix
{\Cal D}^0f(c_1)\\
(c_2-c_1){\Cal D}^1 f(c_1)\\
.\\
.\\
.\\
(c_k-c_1)(c_k-c_2)\dotsc(c_k-c_{k-1}){\Cal D}^{k-1}f(c_1)
\endmatrix\right],\tag2.24
$$
where $\Cal D$ is Newton's divided difference operator, whose powers are defined inductively
by ${\Cal D}^0 f=f$ and 
${\Cal D}^r f(c_j)=({\Cal D}^{r-1} f(c_{j+1})- {\Cal D}^{r-1} f(c_{j}))/(c_{j+r}-c_j)$, $r\geq1$.

This operation has an obvious analog for columns.

As noted in \cite{\ec,\S5} (and as can be seen by looking at (2.11), (2.9), (2.3) and (2.4)), 
operation (2.24) can be applied a total of $2m+2n$ different times to the matrix $M$: each row 
of block matrices in the expression for $M$ given by (2.11) and (2.9) provides two opportunities 
(along the odd-indexed rows and along the even-indexed ones), and each column consisting of
$A$-blocks provides two more. Let $M'$ be the matrix obtained from $M$ after applying these
$2m+2n$ operations.

Since operations (2.24) preserve the determinant (see \cite{\ec,Lemma 5.2}), 
$$
\det M=\det M'.\tag2.25
$$
Moreover, by construction, the $2\times2$
blocks of $M'$ are obtained by replacing $Rx_1\leftarrow x_1^{(R)},\dotsc,Rw_n\leftarrow w_n^{(R)}$ 
in formulas \cite{\ec, (5.6)--(5.9)}. Propositions \secA.1 and \secA.5 and the arguments in the 
proof of \cite{\ec,Lemma 5.3} imply then that\footnote
{\,
For a matrix $A$ whose entries depend on a large parameter, 
$\lfloor\lfloor A\rfloor\rfloor$ stands for the matrix obtained from it by replacing each entry
by the dominant part in its asymptotics as the parameter approaches infinity.
}
$$
\align
\det \lfloor\lfloor M'\rfloor\rfloor = 
\left(\frac{1}{2\pi i}\right)^{2S}\prod_{j=1}^m \ &\prod_{1\leq k<l\leq s_j} (a_{jk}-a_{jl})^2
\prod_{j=1}^n \ \prod_{1\leq k<l\leq t_j} (b_{jk}-b_{jl})^2 
\\
&
\times
\det(M'')\,\,
R^{2\{\sum_{1\leq k<l\leq m}s_ks_l+\sum_{1\leq k<l\leq n}t_kt_l-\sum_{k=1}^m\sum_{l=1}^ns_kt_l \}},
\tag2.26
\endalign
$$
where ${\bold a}_k=(a_{k1},\dotsc,a_{k,s_k})$ and ${\bold b}_l=(b_{l1},\dotsc,b_{l,t_l})$ for all $k$ 
and $l$.

By (2.25) and (2.26) we get
$$
\spreadlines{3\jot}
\align
\det M=
\left(\frac{1}{2\pi i}\right)^{2S}\prod_{j=1}^m \ &\prod_{1\leq k<l\leq s_j} (a_{jk}-a_{jl})^2
\prod_{j=1}^n \ \prod_{1\leq k<l\leq t_j} (b_{jk}-b_{jl})^2 
\\
&
\times
\det(M'')\,\,
R^{2\{\sum_{1\leq k<l\leq m}s_ks_l+\sum_{1\leq k<l\leq n}t_kt_l-\sum_{k=1}^m\sum_{l=1}^ns_kt_l \}}
\\
&
+o\left(
R^{2\{\sum_{1\leq k<l\leq m}s_ks_l+\sum_{1\leq k<l\leq n}t_kt_l-\sum_{k=1}^m\sum_{l=1}^ns_kt_l\}}
\right).\tag2.27
\endalign
$$
Propositions \secA.1 and \secA.5 imply that the implicit constant above  is uniform in 
$x_1,\dotsc,w_n$.

Next we turn to the asymptotics of $\det M_1$. Note that all the $2m+2n$ 
operations of type (2.24) we applied to $M$ are also well-defined as operations on $M_1$. Indeed, 
it is apparent from (2.9) and (2.3)--(2.6) that each of the $2m+2n$ times we applied (2.24), the 
extensions to $M_1$ of the involved rows or columns of $M$ are also of the form required by this 
operation. 
Let $M_1'$ be the matrix obtained from $M_1$ after applying these $2m+2n$ operations.

A calculation similar to the one that gave (2.26) yields
$$
\spreadlines{3\jot}
\align
&
\det \lfloor\lfloor M_1'\rfloor\rfloor = 
\left(\frac{1}{2\pi i}\right)^{2S+1}\prod_{j=1}^m \ \prod_{1\leq k<l\leq s_j} (a_{jk}-a_{jl})^2
\prod_{j=1}^n \ \prod_{1\leq k<l\leq t_j} (b_{jk}-b_{jl})^2 
\\
&\ \ \ \ \ \ \ \ \ \ \ \ \ \ \ \ 
\times
\det(M_1'')\,\,
R^{2\{\sum_{1\leq k<l\leq m}s_ks_l+\sum_{1\leq k<l\leq n}t_kt_l-\sum_{k=1}^m\sum_{l=1}^ns_kt_l \}-1}.
\tag2.28
\endalign
$$
By the preservation of the determinant we get
$$
\spreadlines{3\jot}
\align
\det M_1=
\left(\frac{1}{2\pi i}\right)^{2S+1}\prod_{j=1}^m \ &\prod_{1\leq k<l\leq s_j} (a_{jk}-a_{jl})^2
\prod_{j=1}^n \ \prod_{1\leq k<l\leq t_j} (b_{jk}-b_{jl})^2 
\\
&
\times
\det(M_1'')\,\,
R^{2\{\sum_{1\leq k<l\leq m}s_ks_l+\sum_{1\leq k<l\leq n}t_kt_l-\sum_{k=1}^m\sum_{l=1}^ns_kt_l \}-1}
\\
&
+o\left(R^{2\{\sum_{1\leq k<l\leq m}s_ks_l+\sum_{1\leq k<l\leq n}t_kt_l-\sum_{k=1}^m\sum_{l=1}^ns_kt_l\} 
-1}\right),\tag2.29
\endalign
$$
and the implicit constant is again uniform in $x_1,\dotsc,w_n$ by Propositions \secA.1 and \secA.5.
The statement of the Proposition follows now by (2.22), (2.27) and (2.29). \epf

The same analysis proves also the following result.

\proclaim{Proposition 2.3} If $x_0^{(R)}/R\to x_0,\dotsc,w_n^{(R)}/R\to w_n$ and
$x_0^{(R)}=\alpha_0\,(\text{\rm mod}\ 3),\dotsc,w_n^{(R)}=\delta_n\,(\text{\rm mod}\,\, 3)$, 
we have
$$
\spreadlines{3\jot}
\align
p_2:=&\frac{
\hat\omega\left(L_2\left(x_0^{(R)},y_0^{(R)}\right),
E_{{\bold a}_1}^q\left(x_1^{(R)},y_1^{(R)}\right),\dotsc,
W_{{\bold b}_n}^q\left(z_n^{(R)},w_n^{(R)}\right)\right)
}
{
\hat\omega\left(
E_{{\bold a}_1}^q\left(x_1^{(R)},y_1^{(R)}\right),\dotsc,
W_{{\bold b}_n}^q\left(z_n^{(R)},w_n^{(R)}\right)\right)
}
\\
=&
\frac{1}{3}
+\frac{1}{2\pi i R}\ \frac{\det M_2''}{\det M''}
+o\left(\frac{1}{R}\right),\tag2.30
\endalign
$$
where 
$$
\spreadlines{4\jot}
\align
&
M_2''=
\\
&\ 
\left[\!\matrix
0\!\!&&
\ddot{C}_{x_0,y_0,z_1,w_1}^{\alpha_0,\beta_0,\gamma_1,\delta_1}(t_1)
\!\!&&
\cdots\!\!&&
\ddot{C}_{x_0,y_0,z_n,w_n}^{\alpha_0,\beta_0,\gamma_n,\delta_n}(t_n)
\!\!&&
\ddot{G}_{x_0,y_0}^{\alpha_0,\beta_0}(S-T-1)
\\
\\
\\
\dot{D}_{x_1,y_1,x_0,y_0}^{\alpha_1,\beta_1,\alpha_0,\beta_0}(s_1)\!\!&&
\dot{A}_{x_1,y_1,z_1,w_1}^{\alpha_1,\beta_1,\gamma_1,\delta_1}(s_1,t_1)\!\!&&\cdots\!\!&&
\dot{A}_{x_1,y_1,z_n,w_n}^{\alpha_1,\beta_1,\gamma_n,\delta_n}(s_1,t_n)\!\!&&
\dot{B}_{x_1,y_1}^{\alpha_1,\beta_1}(s_1,S-T-1)
\\
\\
\\
\vdots\!\!&&\vdots\!\!&&\ \!\!&&\vdots\!\!&&\vdots
\\
\\
\\
\dot{D}_{x_m,y_m,x_0,y_0}^{\alpha_m,\beta_m,\alpha_0,\beta_0}(s_m)\!\!&&
\dot{A}_{x_m,y_m,z_1,w_1}^{\alpha_m,\beta_m,\gamma_1,\delta_1}(s_m,t_1)\!\!&&
\cdots\!\!&&
\dot{A}_{x_m,y_m,z_n,w_n}^{\alpha_m,\beta_m,\gamma_n,\delta_n}(s_m,t_n)\!\!&&
\dot{B}_{x_m,y_m}^{\alpha_m,\beta_m}(s_m,S-T-1)
\endmatrix\!\right]\!,
\\
\tag2.31
\endalign
$$
$$
\ddot{C}_{x,y,z,w}^{\alpha,\beta,\gamma,\delta}(t)=
\left[\matrix
{ \cdots}&&
{ \left\langle\frac{ \z^{-1+(\alpha-\beta)-(\gamma-\delta)}(1-q\z)^{j-1} }
{ [z-x-(w-y)\z]^{j} }\right\rangle }
\!\!\!\!\!\!\!\!&&
{ \left\langle\frac{\z^{-3+(\alpha-\beta)-(\gamma-\delta)}(1-q\z)^{j-1}}
{[z-x-(w-y)\z]^{j}}\right\rangle}
&&{ \cdots}
\endmatrix\right],
\tag2.32
$$
and
$$
\ddot{G}_{x,y}^{\alpha,\beta}(k)=
\left[\matrix
\left\langle\frac{\z^{-1+\alpha-\beta}}{(x-y\z)^0}\right\rangle
\!\!\!\!\!\!\!\!\!\!&&
\left\langle\frac{\z^{-3+\alpha-\beta}}{(x-y\z)^0}\right\rangle
\!\!\!\!\!\!\!\!\!\!&&
\left\langle\frac{\z^{-1+\alpha-\beta}}{(x-y\z)^{-1}}\right\rangle
\!\!\!\!\!\!\!\!\!\!&&
\left\langle\frac{\z^{-3+\alpha-\beta}}{(x-y\z)^{-1}}\right\rangle
\!\!\!\!\!\!\!\!\!\!&&
{ \cdots}\!\!\!\!\!\!\!\!\!\!&&
\left\langle\frac{\z^{-1+\alpha-\beta}}{(x-y\z)^{-k}}\right\rangle
\!\!\!\!\!\!\!\!\!\!&&
\left\langle\frac{\z^{-3+\alpha-\beta}}{(x-y\z)^{-k}}\right\rangle
\endmatrix\right].
\tag2.33
$$
The implicit constant above has the same uniformity property as in Proposition $2.2$.

\endproclaim

\mysec{3. Evaluation of the determinants and proof of Theorem 1.1}

The orthogonal projections of the vectors ${\bold e}_1$, ${\bold e}_2$ and ${\bold e}_3$ of (1.1) on
the $x$-axis of our oblique coordinate system are $\sqrt{3}/2$, $-\sqrt{3}/2$ and 0, respectively.
Thus the orthogonal projection of ${\bold F}$ on the $x$-axis is
$$
F_x=\frac{\sqrt{3}}{2}(p_1-p_2).\tag3.1
$$
By (2.13) and (2.30), the part about $F_x$ of Theorem 1.1 will follow provided we show that
$(\det M_1''-\det M_2'')/\det M''$ evaluates to the expression that together with (3.1), (2.13) and (2.30) 
implies (1.2).

\proclaim{Proposition 3.1} 
The determinants of the matrices $M_1''$ given by $(2.14)$ and 
$(2.16)$--$(2.20)$, $M_2''$ given by  $(2.31)$--$(2.33)$,  
and $M''$ given by $(2.15)$, satisfy
$$
\spreadlines{3\jot}
\align
&
\frac{\det M_1''-\det M_2''}{\det M''}=
i\sqrt{3}
\left\{
\sum_{k=1}^m s_k
\frac{2(x_0-x_k)+y_0-y_k}{(x_0-x_k)^2+(x_0-x_k)(y_0-y_k)+(y_0-y_k)^2}
\right.
\\
&\ \ \ \ \ \ \ \ \ \ \ \ \ \ \ \ \ \ \ \ 
\left.
-\sum_{l=1}^n t_l
\frac{2(x_0-z_l)+y_0-w_l}{(x_0-z_l)^2+(x_0-z_l)(y_0-w_l)+(y_0-w_l)^2}
\right\}.\tag3.2
\endalign
$$

\endproclaim

A quick glance at (2.14) and (2.31) shows that $M_1''$ and $M_2''$ differ only in their first rows.
Furthermore, by (2.18), (2.20), (2.32) and (2.33), the corresponding first row entries differ just by a factor of
$\zeta^{-1}$. This makes it very tempting to write the numerator on the left hand side of (3.2) as
a single determinant, and try to use the method of factor exhaustion (after all, this was the method
that proved successful in \cite{ec}, where an explicit product expression for a special case of $\det M''$ 
is given; see \cite{\ec,Theorem 8.1}). 
However, we were not able to prove (3.2) this way, despite
getting frustratingly close (the only missing part was proving divisibility by one last type of
linear polynomial divisor). 

The proof below derives (3.2) by a certain limiting process
from a convenient specialization of \cite{\ec,The\-o\-rem 8.1}. This has the advantage of being
significantly shorter than a possible proof by factor exhaustion. Still, a solution of the latter kind
would be interesting, as it should prove also a conjectured two-parameter 
generalization 
of (1.2) that works in particular with $\z$ being an indeterminate. 

\pf We first prove the statement under the simplifying assumption that $\alpha_0=\cdots=\delta_n=0$.
In this case
Theorem 8.1 of \cite{\ec} implies that for any indeterminates $x_1,\dotsc,x_m$, $y_1,\dotsc,y_m$, 
$z_1,\dotsc,z_n$ and $w_1,\dotsc,w_n$ we have 
$$
\det M''=\det\left[\matrix
\dot{A}_{x_1,y_1,z_1,w_1}^{0,0,0,0}(s_1,t_1)\!\!&&\cdots\!\!&&
\dot{A}_{x_1,y_1,z_n,w_n}^{0,0,0,0}(s_1,t_n)\!\!&&
\dot{B}_{x_1,y_1}^{0,0}(s_1,S-T-1)
\\
\\
\\
\vdots\!\!&&\  \!\!&&\vdots\!\!&&\vdots
\\
\\
\\
\dot{A}_{x_m,y_m,z_1,w_1}^{0,0,0,0}(s_m,t_1)\!\!&&\cdots\!\!&&
\dot{A}_{x_m,y_m,z_n,w_n}^{0,0,0,0}(s_m,t_n)\!\!&&
\dot{B}_{x_m,y_m}^{0,0}(s_m,S-T-1)
\endmatrix\right]
$$
\vskip-0.2in
$$
\spreadlines{4\jot}
\align
&
=
(-3)^S
[(q-\zeta)(q-\zeta^{-1})]^{\sum_{k=1}^m {s_k \choose 2}+\sum_{l=1}^n {t_l \choose 2}}
\\
&
\times
\frac
{
{\displaystyle \prod_{1\leq k<l\leq m}}
{\scriptstyle \left[(x_k-x_l-\zeta(y_k-y_l))
\left(x_k-x_l-\frac{y_k-y_l}{\z}\right)\right]^{s_ks_l}}
{\displaystyle \prod_{1\leq k<l\leq n}}
{\scriptstyle \left[(z_k-z_l-\zeta(w_k-w_l))\left(z_k-z_l-\frac{w_k-w_l}{\z}\right)\right]^{t_kt_l}}
}
{
{\displaystyle \prod_{k=1}^m\prod_{l=1}^n}
{\scriptstyle \left[(x_k-z_l-\zeta(y_k-w_l))\left(x_k-z_l-\frac{y_k-w_l}{\z}\right)\right]^{s_kt_l}}
}.
\tag3.3
\endalign
$$
Let $\epsilon>0$.
Note that the matrix in (3.3) depends on four lists of indeterminates --- $x_1,\dotsc,x_m$, 
$y_1,\dotsc,y_m$, $z_1,\dotsc,z_n$, and $w_1,\dotsc,w_n$ --- and two lists of positive integers
$s_1,\dotsc,s_m$ and $t_1,\dotsc,t_n$.    
Replace the indeterminates as follows: the $x$-list by $x_0,x_1,\dotsc,x_m$, 
the $y$-list by $y_0,y_1,\dotsc,y_m$, the
$z$-list by $x_0+\epsilon,z_1,\dotsc,z_n$, and the $w$-list by $y_0,w_1,\dotsc,w_n$.
Replace also the $s$-list by $2,s_1,s_2,\dotsc,s_m$, and the $t$-list by $2,t_1,t_2,\dotsc,t_n$.
Then (3.3) provides 
an explicit product expression for the determinant of the matrix
$$
\spreadlines{3\jot}
\align
&
M_\epsilon=
\\
&
\left[\!\matrix
\left\langle\frac{\z^{-1}}{\epsilon}\right\rangle\!\!\!\!\!\!\!\!&&
\left\langle\frac{\z^{-3}}{\epsilon}\right\rangle\!\!\!\!\!\!\!\!&&
\ddot{C}_{x_0,y_0,z_1,w_1}^{0,0,0,0}(t_1)\!\!\!\!\!\!\!\!&&
\cdots\!\!\!\!\!\!\!\!&&
\ddot{C}_{x_0,y_0,z_n,w_n}^{0,0,0,0}(t_n)\!\!\!\!\!\!\!\!&&
\ddot{G}_{x_0,y_0}^{0,0}(\nu)
\\
\\
\left\langle\frac{\z}{\epsilon}\right\rangle\!\!\!\!\!\!\!\!&&
\left\langle\frac{\z^{-1}}{\epsilon}\right\rangle\!\!\!\!\!\!\!\!&&
\dddot{C}_{x_0,y_0,z_1,w_1}(t_1)\!\!\!\!\!\!\!\!&&
\cdots\!\!\!\!\!\!\!\!&&
\dddot{C}_{x_0,y_0,z_n,w_n}(t_n)\!\!\!\!\!\!\!\!&&
\dddot{G}_{x_0,y_0}(\nu)
\\
\\
\dddot{D}_{x_1,y_1,x_0+\epsilon,y_0}(s_1)\!\!\!\!\!\!\!\!&&
\ddot{D}_{x_1,y_1,x_0+\epsilon,y_0}(s_1)\!\!\!\!\!\!\!\!&&
\dot{A}_{x_1,y_1,z_1,w_1}^{0,0,0,0}(s_1,t_1)\!\!\!\!\!\!\!\!&&
\cdots\!\!\!\!\!\!\!\!&&
\dot{A}_{x_1,y_1,z_n,w_n}^{0,0,0,0}(s_1,t_n)\!\!\!\!\!\!\!\!&&
\dot{B}_{x_1,y_1}^{0,0}(s_1,\nu)
\\
\\
\vdots\!\!\!\!\!\!\!\!&&\vdots\!\!\!\!\!\!\!\!&&\vdots\!\!\!\!\!\!\!\!&&\ \!\!\!\!\!\!\!\!&&
\vdots\!\!\!\!\!\!\!\!&&\vdots
\\
\\
\dddot{D}_{x_m,y_m,x_0+\epsilon,y_0}(s_1)\!\!\!\!\!\!\!\!&&
\ddot{D}_{x_m,y_m,x_0+\epsilon,y_0}(s_1)\!\!\!\!\!\!\!\!&&
\dot{A}_{x_m,y_m,z_1,w_1}^{0,0,0,0}(s_m,t_1)\!\!\!\!\!\!\!\!&&
\cdots\!\!\!\!\!\!\!\!&&
\dot{A}_{x_m,y_m,z_n,w_n}^{0,0,0,0}(s_m,t_n)\!\!\!\!\!\!\!\!&&
\dot{B}_{x_m,y_m}^{0,0}(s_m,\nu)
\endmatrix\!\!\right]\!\!,
\\
\tag3.4
\endalign
$$
where $\nu=S-T-1$,
$$
\dddot{C}_{x,y,z,w}(t)=
\left[\matrix
{ \cdots}&&
{ \left\langle\frac{ \z(1-q\z)^{j-1} }{ [z-x-(w-y)\z]^{j} }\right\rangle }
\!\!\!\!\!\!\!\!&&
{ \left\langle\frac{\z^{-1}(1-q\z)^{j-1}}{[z-x-(w-y)\z]^{j}}\right\rangle}
&&{ \cdots}
\endmatrix\right],
\tag3.5
$$

$$
\spreadmatrixlines{1\jot}
\ddot{D}_{x,y,z,w}(s)=
\left[\matrix
\left\langle\frac{\z^{-3}(1-q\z)^0}{[z-x-(w-y)\z]}\right\rangle\\
\left\langle\frac{\z^{-1}(1-q\z)^0}{[z-x-(w-y)\z]}\right\rangle\\
\\
\left\langle\frac{\z^{-3}(1-q\z)^1}{[z-x-(w-y)\z]^2}\right\rangle\\
\left\langle\frac{\z^{-1}(1-q\z)^1}{[z-x-(w-y)\z]^2}\right\rangle\\
\\
\vdots\\
\\
\left\langle\frac{\z^{-3}(1-q\z)^{s-1}}{[z-x-(w-y)\z]^s}\right\rangle\\
\left\langle\frac{\z^{-1}(1-q\z)^{s-1}}{[z-x-(w-y)\z]^s}\right\rangle
\endmatrix\right],
\ \ \ (3.6)\ \ \ \ \ \  
\dddot{D}_{x,y,z,w}(s)=
\left[\matrix
\left\langle\frac{\z^{-1}(1-q\z)^0}{[z-x-(w-y)\z]}\right\rangle\\
\left\langle\frac{\z(1-q\z)^0}{[z-x-(w-y)\z]}\right\rangle\\
\\
\left\langle\frac{\z^{-1}(1-q\z)^1}{[z-x-(w-y)\z]^2}\right\rangle\\
\left\langle\frac{\z(1-q\z)^1}{[z-x-(w-y)\z]^2}\right\rangle\\
\\
\vdots\\
\\
\left\langle\frac{\z^{-1}(1-q\z)^{s-1}}{[z-x-(w-y)\z]^s}\right\rangle\\
\left\langle\frac{\z(1-q\z)^{s-1}}{[z-x-(w-y)\z]^s}\right\rangle
\endmatrix\right],
\tag3.7
$$
and
$$
\dddot{G}_{x,y}(k)=
\left[\matrix
\left\langle\frac{\z}{(x-y\z)^0}\right\rangle
\!\!\!\!\!\!\!\!\!\!&&
\left\langle\frac{\z^{-1}}{(x-y\z)^0}\right\rangle
\!\!\!\!\!\!\!\!\!\!&&
\left\langle\frac{\z}{(x-y\z)^{-1}}\right\rangle
\!\!\!\!\!\!\!\!\!\!&&
\left\langle\frac{\z^{-1}}{(x-y\z)^{-1}}\right\rangle
\!\!\!\!\!\!\!\!\!\!&&
{ \cdots}\!\!\!\!\!\!\!\!\!\!&&
\left\langle\frac{\z}{(x-y\z)^{-k}}\right\rangle
\!\!\!\!\!\!\!\!\!\!&&
\left\langle\frac{\z^{-1}}{(x-y\z)^{-k}}\right\rangle
\endmatrix\right].
\tag3.8
$$
The resulting product expression for $\det M_\epsilon$
has many common factors with the right hand side of (3.3), which
simplify when taking $\det M_\epsilon/\det M''$. After simplification we get
$$
\align
&
\frac{\det M_\epsilon}{\det M''}=
\\
&\ \ 
-\frac{3}{\epsilon^2}
\ \frac
{
{\displaystyle \prod_{k=1}^m}
{\scriptstyle \left[\left(x_0-x_k-\z(y_0-y_k)\right)\left(x_0-x_k-\frac{y_0-y_k}{\z}\right)\right]^{s_k}}
{\displaystyle \prod_{l=1}^n}
{\scriptstyle \left[\left(x_0+\epsilon-z_l-\z(y_0-w_l)\right)
\left(x_0+\epsilon-z_l-\frac{y_0-w_l}{\z}\right)\right]^{t_l}}
}
{
{\displaystyle \prod_{k=1}^m}
{\scriptstyle \left[\left(x_k-x_0-\epsilon-\z(y_k-y_0)\right)
\left(x_k-x_0-\epsilon-\frac{y_k-y_0}{\z}\right)\right]^{s_k}}
{\displaystyle \prod_{l=1}^n}
{\scriptstyle \left[\left(x_0-z_l-\z(y_0-w_l)\right)
\left(x_0-z_l-\frac{y_0-w_l}{\z}\right)\right]^{t_l}}
}.
\\
\tag3.9
\endalign
$$
Replace the first column in (3.4) by the negative of the sum of the first two columns. Using
$\z^3=1$, $-\z^{-1}-\z^{-3}=\z^{-2}$, and $-\z-\z^{-1}=1$, we obtain
$$
\spreadlines{3\jot}
\spreadmatrixlines{1\jot}
\align
&
-\det M_\epsilon=
\\
&
\left|\!\matrix
\frac{\z-\z^{-1}}{\epsilon}\!\!\!\!\!\!\!\!&&
0\!\!\!\!\!\!\!\!&&
\ddot{C}_{x_0,y_0,z_1,w_1}^{0,0,0,0}(t_1)\!\!\!\!\!\!\!\!&&
\cdots\!\!\!\!\!\!\!\!&&
\ddot{C}_{x_0,y_0,z_n,w_n}^{0,0,0,0}(t_n)\!\!\!\!\!\!\!\!&&
\ddot{G}_{x_0,y_0}^{0,0}(\nu)
\\
\\
0\!\!\!\!\!\!\!\!&&
\frac{\z^{-1}-\z}{\epsilon}\!\!\!\!\!\!\!\!&&
\dddot{C}_{x_0,y_0,z_1,w_1}(t_1)\!\!\!\!\!\!\!\!&&
\cdots\!\!\!\!\!\!\!\!&&
\dddot{C}_{x_0,y_0,z_n,w_n}(t_n)\!\!\!\!\!\!\!\!&&
\dddot{G}_{x_0,y_0}(\nu)
\\
\\
\dot{D}_{x_1,y_1,x_0+\epsilon,y_0}^{0,0,0,0}(s_1)\!\!\!\!\!\!\!\!&&
\ddot{D}_{x_1,y_1,x_0+\epsilon,y_0}(s_1)\!\!\!\!\!\!\!\!&&
\dot{A}_{x_1,y_1,z_1,w_1}^{0,0,0,0}(s_1,t_1)\!\!\!\!\!\!\!\!&&
\cdots\!\!\!\!\!\!\!\!&&
\dot{A}_{x_1,y_1,z_n,w_n}^{0,0,0,0}(s_1,t_n)\!\!\!\!\!\!\!\!&&
\dot{B}_{x_1,y_1}^{0,0}(s_1,\nu)
\\
\\
\vdots\!\!\!\!\!\!\!\!&&\vdots\!\!\!\!\!\!\!\!&&\vdots\!\!\!\!\!\!\!\!&&\ \!\!\!\!\!\!\!\!&&
\vdots\!\!\!\!\!\!\!\!&&\vdots
\\
\\
\dot{D}_{x_m,y_m,x_0+\epsilon,y_0}^{0,0,0,0}(s_1)\!\!\!\!\!\!\!\!&&
\ddot{D}_{x_m,y_m,x_0+\epsilon,y_0}(s_1)\!\!\!\!\!\!\!\!&&
\dot{A}_{x_m,y_m,z_1,w_1}^{0,0,0,0}(s_m,t_1)\!\!\!\!\!\!\!\!&&
\cdots\!\!\!\!\!\!\!\!&&
\dot{A}_{x_m,y_m,z_n,w_n}^{0,0,0,0}(s_m,t_n)\!\!\!\!\!\!\!\!&&
\dot{B}_{x_m,y_m}^{0,0}(s_m,\nu)
\endmatrix\!\right|\!.
\\
\tag3.10
\endalign
$$
Denote the matrix on the right hand side above by $N=(N_{ij})_{1\leq i,j\leq 2S+1}$. As $\epsilon\to0$,
the only diverging entries of $N$ are $N_{11}$ and $N_{22}$. It follows from (3.10) that
$$
\spreadlines{4\jot}
\align
&
\det M_\epsilon=\frac{(\z-\z^{-1})^2}{\epsilon^2}
\det N_{[2S+1]\setminus\{1,2\}}^{[2S+1]\setminus\{1,2\}}
+\frac{\z^{-1}-\z}{\epsilon}
\left[\det \bar{N}_{[2S+1]\setminus\{1\}}^{[2S+1]\setminus\{1\}}
-\det \bar{N}_{[2S+1]\setminus\{2\}}^{[2S+1]\setminus\{2\}}\right]
\\
&\ \ \ \ \ \ \ \ \ \ \ \ \ \ \ \ \ \ \ \ \ \ \ \ \ \ \ \ \ \ \ \ 
\ \ \ \ \ \ \ \ \ \ \ \ \ \ \ \ \ \ \ \ \
+O(1),\ \ \ \epsilon\to 0,\tag3.11
\endalign
$$
where $\bar{N}$ is the matrix obtained from $N$ by replacing $N_{11}$ and $N_{22}$ by 0. But 
$\bar{N}_{[2S+1]\setminus\{2\}}^{[2S+1]\setminus\{2\}}$ is just the matrix $M_2''$ of (2.31),
and $N_{[2S+1]\setminus\{1,2\}}^{[2S+1]\setminus\{1,2\}}=M''$.
Furthermore, by Lemma 3.2, 
$\det \bar{N}_{[2S+1]\setminus\{1\}}^{[2S+1]\setminus\{1\}}=\det M_1''$. Therefore, (3.11) implies
$$
\frac{\det M_\epsilon}{\det M''}=
-\frac{3}{\epsilon^2}+\frac{\z^{-1}-\z}{\epsilon}\ \frac{\det M_1''-\det M_2''}{\det M''}
+O(1),\ \ \ \epsilon\to 0.\tag3.12
$$
Extracting the coefficient of $1/\epsilon$ in the asymptotics of (3.9) as $\epsilon\to 0$
and comparing it with the second term in (3.12) gives (3.2). 

Next we show how the case of general residues $\alpha_0,\dotsc,\delta_n$ modulo 3 reduces to the
above case of all zero residues.

Consider the general matrix $M_1''$ given by (2.14) and (2.16)--(2.20). Except for its first row and column, 
it consists of $2\times2$ blocks of the type in Lemma 3.3. Furthermore, the corresponding
value of $a$ is constant over each block 
$\dot{A}_{x_k,y_k,z_l,w_l}^{\alpha_k,\beta_k,\gamma_l,\delta_l}(s_k,t_l)$, and equals 
$(\alpha_k-\beta_k)-(\gamma_l-\delta_l)$. It is also constant over each block 
$\dot{B}_{x_k,y_k}^{\alpha_k,\beta_k}(s_k,S-T-1)$, and equals $\alpha_k-\beta_k$. Thus in each
$2\times2$ block of type (3.13) in $M_1''$ the value of the $a$ of Lemma 3.3 is a difference of two 
quantities, the first being constant along each row of $M_1''$, and the second constant along each
column of $M_1''$. Therefore Lemma 3.3 can be used to transform the general $M_1''$ matrix 
by row and column operations into the 
specialization of $M_1''$ when all residues are 0 except $\alpha_0$ and $\beta_0$, and the overall 
effect on the determinant is that it is multiplied by $\sigma$, where $\sigma\in\{1,-1\}$.
The very same row and column operations transform $M_2''$ into $M_2''|_{\alpha_1=0,\dotsc,\delta_n=0}$
and $M''$ into $M''|_{\alpha_1=0,\dotsc,\delta_n=0}$, with the effect on their determinants being
multiplication by the same $\sigma$. Therefore
$$
\frac{\det M_1''-\det M_2''}{\det M''}
=
\frac{\det M_1''|_{\alpha_1=0,\dotsc,\delta_n=0}-\det M_2''|_{\alpha_1=0,\dotsc,\delta_n=0}}
{\det M''|_{\alpha_1=0,\dotsc,\delta_n=0}}.
$$
However, Lemma 3.4 provides determinant preserving operations that transform all three determinants
above into their $\alpha_0=0$, $\beta_0=0$ specializations. \epf

\proclaim{Lemma 3.2} $\det \bar{N}_{[2S+1]\setminus\{1\}}^{[2S+1]\setminus\{1\}}=\det M_1''$.
\endproclaim

\pf 
Repeated application of Lemma 3.4 
transforms the first row and column of the matrix $\bar{N}_{[2S+1]\setminus\{1\}}^{[2S+1]\setminus\{1\}}$ 
(see (3.5), (3.7) and (3.8)) into the first row and column of the matrix $M_1''$ (see (2.18)--(2.20)).
All the other entries of $\bar{N}_{[2S+1]\setminus\{1\}}^{[2S+1]\setminus\{1\}}$ and $\det M_1''$ agree,
and are left in agreement by the above applications of Lemma~3.4. This proves the claim. \epf

\proclaim{Lemma 3.3} Let $f$ be a function defined at $\z$ and $\z^{-1}$, and let $a\in\Z$. 
Let
$$
\spreadmatrixlines{1\jot}
A(a)=\left[\matrix
\left\langle\z^{a-1}f(\z)\right\rangle&&\left\langle\z^{a-3}f(\z)\right\rangle
\\
\left\langle\z^{a+1} f(\z)\right\rangle&&\left\langle\z^{a-1}f(\z)\right\rangle
\endmatrix\right].
$$
Denote its rows by $R_1$ and $R_2$, and its columns by $C_1$ and $C_2$.

$(${\rm a}$)$. Simultaneously replacing $\{R_1\leftarrow R_2,R_2\leftarrow -R_1-R_2\}$ 
turns matrix $A(a)$ into $A(a-1)$.

$(${\rm b}$)$. Simultaneously replacing $\{C_2\leftarrow C_1,C_1\leftarrow -C_1-C_2\}$ 
turns matrix $A(a)$ into $A(a+1)$.

\endproclaim

\pf 
Since $\z^3=1$, the second row of $A(a)$ is the same as the first row of $A(a-1)$. 
The negative of the sum of the entries in the first column of $A(a)$ is
$$
-\left(\z^{a-1}f(\z)-\z^{-a+1}f(\z^{-1})\right)
-\left(\z^{a+1}f(\z)-\z^{-a-1}f(\z^{-1})\right)
=
\z^{a}f(\z)-\z^{-a}f(\z^{-1})
=
\left\langle\z^{a}f(\z)\right\rangle,
$$
as $-\z^{k}-\z^{k+2}=\z^{k+1}$ for all integers $k$. One similarly checks that the negative of the 
sum of the entries in the second column of $A(a)$ equals $\left\langle\z^{a-2}f(\z)\right\rangle$.
This proves (a). Part (b) follows analogously. \epf

By a similar calculation one can easily check the following.

\proclaim{Lemma 3.4} Let $f$ be a function defined at $\z$ and $\z^{-1}$, and let 
$\alpha,\beta,\gamma\in\Z$. 
Form the matrix
$$
\spreadmatrixlines{1\jot}
\left[\matrix
0&&\left\langle\z^{1+\alpha}f(\z)\right\rangle&&\left\langle\z^{-1+\alpha}f(\z)\right\rangle
\\
\left\langle\z^{-3+\beta} f(\z)\right\rangle&&
\left\langle\z^{-1+\gamma} f(\z)\right\rangle&&\left\langle\z^{-3+\gamma}f(\z)\right\rangle
\\
\left\langle\z^{-1+\beta} f(\z)\right\rangle&&
\left\langle\z^{1+\gamma} f(\z)\right\rangle&&\left\langle\z^{-1+\gamma}f(\z)\right\rangle
\endmatrix\right].
$$
Denote its rows by $R_1,R_2,R_3$, its columns by $C_1,C_2,C_3$.
Then the result of the simultaneous column operations $\{C_2\leftarrow -C_2-C_3,C_3\leftarrow C_2\}$,
followed by the simultaneous row operations $\{R_2\leftarrow -R_2-R_3,R_3\leftarrow R_2\}$, is
the matrix
$$
\spreadmatrixlines{1\jot}
\left[\matrix
0&&\left\langle\z^{\alpha}f(\z)\right\rangle&&\left\langle\z^{-2+\alpha}f(\z)\right\rangle
\\
\left\langle\z^{-2+\beta} f(\z)\right\rangle&&
\left\langle\z^{-1+\gamma} f(\z)\right\rangle&&\left\langle\z^{-3+\gamma}f(\z)\right\rangle
\\
\left\langle\z^{\beta} f(\z)\right\rangle&&
\left\langle\z^{1+\gamma} f(\z)\right\rangle&&\left\langle\z^{-1+\gamma}f(\z)\right\rangle
\endmatrix\right]. 
$$

\endproclaim

\medskip
{\it Proof of Theorem 1.1.} Part (1.2) of the statement follows from (3.1), (2.13) and (3.2). Interchanging
the roles of the coordinate axes we get (1.3). \epf

%
%
%
%


\mysec{\secA. A finer integral asymptotics}

Let $\Cal D$ denote Newton's divided difference operator, whose powers are defined inductively by 
$\Cal D^0 f=f$ and $\Cal D^r f(c_j)=(\Cal D^{r-1} f(c_{j+1})-\Cal D^{r-1} f(c_j))/(c_{j+r}-c_j)$, $r\geq1$.
We will need the following result on the asymptotics of the coupling function $P$ when acted on in the
indicated way by powers of $\Cal D$. 

\proclaim{Theorem \secA.1} Let $r_n$ and $s_n$ be integers so that $\lim_{n\to\infty}r_n/n=u$, 
$\lim_{n\to\infty}s_n/n=v$, and $(u,v)\neq(0,0)$. Then for any integers $k,l\geq0$ and any rational
number $q$ with $3|1-q$ we have
$$
\align
&
\left.\Cal D^l_y\left\{\Cal D^k_x \,P(r_n+x+y,s_n+q(x+y))|_{x=a_1}\right\}\right|_{y=b_1}=
\\
&\ \ \ \ \ \ \ \ \ \ \ \ \ \ \ \ \ \ \ \ 
\frac{1}{2\pi i}{k+l\choose k}
\left\langle\frac{\zeta^{r_n-s_n-1}(1-q\zeta)^{k+l}}{(-r_n+s_n\zeta)^{k+l+1}}\right\rangle
+O\left(\frac{1}{n^{k+l+2}}\right),\tag\secA.1
\endalign
$$
where $\zeta=e^{2\pi i/3}$, $\Cal D^k_x$ acts with respect to a fixed integer sequence 
$a_1,a_2,\dotsc$, and
$\Cal D^l_y$ acts with respect to an integer sequence $b_1,b_2,\dotsc$ satisfying $qb_j\in\Z$ for all
$j\geq1$. 
Furthermore, for any open set $U$ containing the origin the implicit constant above is uniform for all 
$(u,v)\notin U$. 

\endproclaim

We note that Proposition 7.1 of \cite{\ec} corresponds to the special case when  
$r_n=-un$ and $s_n=-vn$. It will be crucial in our proof of the scaling limit of the average lifting
surface to know that the leading term on the right hand side of (\secA.1) is independent of the way
$r_n/n$ and $s_n/n$ approach their limits.

We deduce Theorem \secA.1 from the following auxiliary results. Let $\Cal P$ denote the 
counterclockwise
oriented arc of the unit circle connecting $\zeta=e^{2\pi i/3}$ to $-1$; $\Cal P[\zeta,-1)$ stands for
this path minus the point $-1$.

\proclaim{Lemma \secA.2} Let $r_n$ and $s_n$ be integers, and assume 
$\lim_{n\to\infty}r_n/n=u \geq \epsilon>0$. Then
if $q\in C^0(\Cal P[\zeta,-1))$ has a pole of finite order at $-1$, for any integer
$k\geq0$ we have
$$
\int_{\zeta}^{-1}(-1-t)^{r_n}t^{s_n}(t-\zeta)^kq(t)\,dt=O\left(\frac{1}{n^{k+1}}\right),\tag\secA.2
$$
where the implicit constant depends only on $k$, $\epsilon$ and $q(t)$ $($in particular, it works for
all $u\geq\epsilon$ and all integers $s_n$$)$.

\endproclaim

\pf Let $I$ denote the integral above. Write $q(t)=\tilde{q}(t)/(t+1)^l$, with $l\geq0$ and 
$\tilde{q}\in C^0(\Cal P)$. Making the change of variable $t=e^{i\theta}$ we obtain
$$
\align
I&=(-1)^{r_n}\int_{\zeta}^{-1}(1+t)^{r_n-l}t^{s_n}(t-\zeta)^k\tilde{q}(t)\,dt
\\
&=(-1)^{r_n}\int_{2\pi/3}^{\pi}\left(2\cos\frac\theta 2 \,e^{i\theta/2} \right)^{r_n-l} 
e^{i\theta s_n}
\left(e^{i\theta}-e^{2\pi i/3}\right)^k\tilde{q}\left(e^{i\theta}\right)i e^{i\theta}\,d\theta
\\
&=(-1)^{r_n}\int_{0}^{\pi/3}\left[2\cos\left(\frac{\pi}{3}+\frac{\tau}{2}\right) \,
e^{i(\pi/3+\tau/2)} \right]^{r_n-l} 
e^{i(2\pi/3+\tau) s_n}
\\
&\ \ \ \ \ \ \ \ \ \ \ \ \ \ \ \ \ \ \ \ 
\times\left[e^{2\pi i/3}\left(e^{i\tau}-1\right)\right]^k
\tilde{q}\left(e^{i(2\pi/3+\tau)}\right)i e^{i(2\pi/3+\tau)}\,d\tau.\tag\secA.3
\endalign
$$
Since the graph of $2\cos(\pi/3+\tau/2)$ is concave for $\tau\in[0,\pi/3]$, it lies below its tangent
at $\tau=0$. This implies
$$
2\cos\left(\frac{\pi}{3}+\frac{\tau}{2}\right)\leq1-\frac{\sqrt{3}}{2}\tau<1-\frac\tau2, 
\ \ \ 0\leq\tau\leq\pi/3.
$$
The elementary inequality $|e^y-1|\leq|y|e^{|y|}$ implies
$$
|e^{i\tau}-1|\leq \tau e^\tau <3\tau,\ \ \ \tau\in[0,\pi/3].
$$
The above two inequalities combined with (\secA.3) give
$$
\align
|I|&\leq\int_0^{\pi/3}\left(1-\frac\tau2\right)^{r_n-l}(3\tau)^k
\left|\tilde{q}\left(e^{i(2\pi/3+\tau)}\right)\right|\,d\tau
\\
&\leq 3^k\sup_{\tau\in\Cal P}|\tilde{q}(t)| \int_0^{\pi/3}\left(1-\frac\tau2\right)^{r_n-l}
\tau^k\,d\tau.\tag\secA.4
\endalign
$$
However, integration by parts implies
$$
\align
J(r,k):=&\int_0^{\pi/3}\left(1-\frac\tau2\right)^r \tau^k\,d\tau
\\
=&
-\frac{2}{r+1}\left(1-\frac\pi6\right)^{r+1}\left(\frac\pi3\right)^k+\frac{2k}{r+1}J(r+1,k-1).
\endalign
$$
Repeated application of this, together with
$$
J(r+k,0)=-\frac{2}{r+k+1}\left(1-\frac\pi6\right)^{r+k+1}+\frac{2}{r+k+2}
$$
shows that $J(r,k)$ is equal to a sum of $k+1$ terms each exponentially small in $r$, plus 
$$
\frac{2}{r+k+2}\frac{(2k)(2k-2)\cdots2}{(r+1)(r+2)\cdots(r+k)}.
$$
Since $r_n-l+1,\dotsc,r_n-l+k+2\geq \frac{1}{2}nu$ for $n$ large enough,
it follows that the integral in the second line of (\secA.4) is 
majorized for $n$ large enough by $2^{2k+3}k!/(nu)^{k+1}\leq 2^{2k+3}k!/(\epsilon)^{k+1} 1/n^{k+1}$, and the proof is 
complete. \epf

\proclaim{Lemma \secA.3} Let $q(t)=(t-\zeta)^kq_1(t)$, where $q_1(t)\in C^1(\Cal P)$, $k$ is a 
non-negative integer, and $q_1(\zeta)\neq0$. Let $r_n,s_n\in\Z$ so that  $\lim_{n\to\infty}r_n/n=u>0$.

$($a$)$. If $k=0$, 
$$
\int_{\zeta}^{-1}(-1-t)^{r_n}t^{s_n}q(t)\,dt=
\frac{\zeta^{s_n-r_n}q(\zeta)}{r_n\zeta-s_n\zeta^{-1}}+O\left(\frac{1}{n^2}\right).\tag\secA.5
$$

$($b$)$. If $k\geq1$, 
$$
\int_{\zeta}^{-1}(-1-t)^{r_n}t^{s_n}q(t)\,dt=
\frac{1}{r_n\zeta-s_n\zeta^{-1}}\int_{\zeta}^{-1}(-1-t)^{r_n}t^{s_n}q'(t)\,dt
+O\left(\frac{1}{n^{k+2}}\right).\tag\secA.6
$$

Furthermore, for any $\epsilon>0$, each implicit constant above can be chosen to be 
uniform for  $u\in[\epsilon,\infty)$ and independent of $s_n$. 

\endproclaim

\pf For $l=2,3,\dotsc,7$, define 
$$
h_l(t):=\sum_{i\geq0}\frac{1}{l+6j}(t-\zeta)^{l+6j}, \ \ \ t\in\Cal P[\zeta,-1).\tag\secA.7
$$
One readily checks that the Taylor series expansions of $\ln t$ and $\ln(-1-t)$ around 
$t=\zeta$ can then be written as
$$
\ln t=\ln \zeta +\zeta^{-1}(t-\zeta)-\zeta^{-1}h_4+\zeta^{-1}h_7-\zeta h_2 +\zeta h_5 +h_3-h_6,
$$
and 
$$
\ln(-1-t)=\ln(-1-\zeta)-\zeta(t-\zeta)-\zeta h_4 -\zeta h_7-\zeta^{-1}h_2-\zeta^{-1}h_5-h_3-h_6.
$$
Therefore we have
$$
\spreadlines{2\jot}
\align
&\!\!\!\!\!\!\!\!
r\ln(-1-t)+s\ln t = \ r\ln(-1-\zeta)+s\ln \zeta+(s\zeta^{-1}-r\zeta)(t-\zeta)
\\
&\ \ \ \ \ \ \ \ \ \ 
+(-r\zeta^{-1}-s\zeta)h_2(t)+(-r+s)h_3(t)
+(-r\zeta-s\zeta^{-1})h_4(t)
\\
&\ \ \ \ \ \ \ \ \ \ 
+(-r\zeta^{-1}+s\zeta)h_5(t)
+(-r-s)h_6(t)+(-r\zeta+s\zeta^{-1})h_7(t),\ \ \ t\in\Cal P[\zeta,-1).\tag\secA.8
\endalign
$$
Let $r,s\in\Z$ and denote
$$
I(r,s):=\int_{\zeta}^{-1}(-1-t)^{r}t^{s}q(t)\,dt.
$$
Using (\secA.8) and $\z^3=1$, integration by parts gives
$$
\spreadlines{2\jot}
\align
&\!
I(r,s)=\int_{\zeta}^{-1}e^{r\ln(-1-t)+s\ln t}q(t)\,dt
\\
&
=\zeta^{s-r}\int_{\zeta}^{-1}e^{(s\zeta^{-1}-r\zeta)(t-\zeta)}
e^{(-r\zeta^{-1}-s\zeta)h_2+\cdots+(-r\zeta+s\zeta^{-1})h_7}q(t)\,dt
\\
&
=\zeta^{s-r}
\left\{\left.
\frac{ e^{(s\zeta^{-1}-r\zeta)(t-\zeta)} }{ s\zeta^{-1}-r\zeta }e^{b(t)}q(t)\right|_\zeta^{-1}\right.
\\
&\!\!\!
\left.
-\frac{1}{ s\zeta^{-1}-r\zeta }\int_{\zeta}^{-1}e^{(s\zeta^{-1}-r\zeta)(t-\zeta)}
\left[\left((-r\zeta^{-1}-s\zeta)h_2'(t)+\cdots+(-r\zeta+s\zeta^{-1})h_7'(t)\right)e^{b(t)}q(t)+e^{b(t)}q'(t)\right]
\,dt
\right\},
\endalign
$$
where 
$$
\spreadlines{2\jot}
\align
&
b(t):=(-r\zeta^{-1}-s\zeta)h_2(t)+(-r+s)h_3(t)
+(-r\zeta-s\zeta^{-1})h_4(t)
\\
&\ \ \ \ \ \ 
+(-r\zeta^{-1}+s\zeta)h_5(t)
+(-r-s)h_6(t)+(-r\zeta+s\zeta^{-1})h_7(t).
\endalign
$$

Since $\zeta^{s-r}e^{(s\zeta^{-1}-r\zeta)(t-\zeta)}e^{b(t)}=(-1-t)^rt^s$, we see that
the upper limit in the first term of the expression in the large curly braces above equals 0
whenever $r>k$. 
Thus, for $r>k$ we obtain
$$
\spreadlines{2\jot}
\align
&
I(r,s)=\frac{\zeta^{s-r}q(\zeta)}{r\zeta-s\zeta^{-1}} 
+\zeta^{s-r}\left\{\frac{-r\zeta^{-1}-s\zeta}{r\z-s\z^{-1}}
\int_\z^{-1}e^{(s\z^{-1}-r\z)(t-\z)}e^{b(t)}q(t)h_2'(t)\,dt+\cdots
\right.
\\
&\ \ \ \ \ \ \ \ \ \ \ \ \ \ \ \ \ \ \ \ \ \ \ \ \ \ \ \ \ \ \ \ \ \ \ \ \ \ \ \ \ \ \ \ \ \ \ \  
\left.
+\frac{-r\z+s\z^{-1}}{r\z-s\z^{-1}}
\int_\z^{-1}e^{(s\z^{-1}-r\z)(t-\z)}e^{b(t)}q(t)h_7'(t)\,dt
\right\}
\\
&\ \ \ \ \ \ \ \ \ \ \ \ \ 
+\frac{\z^{s-r}}{r\z-s\z^{-1}}\int_\z^{-1}e^{(s\z^{-1}-r\z)(t-\z)}e^{b(t)}q'(t)\,dt.\tag\secA.9
\endalign
$$
If $|r|\geq|s|$ we have
$$
\left|\frac{-r\zeta^{-1}-s\zeta}{r\z-s\z^{-1}}\right|=\left|\frac{1+s/r\,\z^{-1}}{1-s/r\,\z}\right|
\leq \frac{1+|s/r|}{|\Rep(1-s/r\,\z)|} \leq \frac{2}{1/2}=4.\tag\secA.10
$$
A similar argument shows that the above inequality holds in fact also when $|s|\geq|r|$. All six 
fractions in front of the integrals in the expression in curly braces above are thus seen to be
majorized in absolute value by 4.

Regard $I(r_n,s_n)$ as the sum of the three quantities provided by (\secA.9). To deduce part (a) of 
the Lemma, assume $k=0$. 
All six terms of the second quantity are $O(1/n^2)$ thanks to (\secA.10) and an 
application of Lemma \secA.2 with $k=1$ (which applies since $h_2'(t),\dotsc,h_7'(t)$ are all
of the form $(t-\zeta)g(t)$, where $g\in C^{\infty}(\Cal P[\z,-1))$ has a simple pole at $t=-1$),
followed by an application of Lemma 3.4 with $k=0$ for the resulting integral on the right hand side of (4.6). 
Finally, the third quantity provided by (\secA.9) is also $O(1/n^2)$, due to the fraction in front of 
the integral and another application of Lemma \secA.2 with $k=0$.

For part (b), assume $k\geq1$. Then $q(\z)=0$, and (\secA.9) provides an expression for $I(r_n,s_n)$
as a sum of just two quantities. All six terms in the first quantity are $O(1/n^{k+2})$ due to
(\secA.10) and Lemma \secA.2 applied with $k$ replaced by $k+1$ (as explained in the previous 
paragraph, this unit increment comes about by the presence of the $h_l'(t)$ factors in the 
integrands). This proves (\secA.6).

The uniformity of the implicit constant follows because both the majorant in (\secA.10) and the 
implicit constant in (\secA.2) are uniform. \epf

Repeated application of part (b) of the above lemma and one final application of part (a) yields the
following result.

\proclaim{Proposition \secA.4} Let $q(t)=(t-\z)^k q_1(t)$, where $k\geq0$, 
$q_1\in C^{k+1}(\Cal P[\z,-1))$ has a pole of finite order at $t=-1$, and $q_1(\z)\neq0$.  
Let $r_n,s_n\in\Z$ so that  $\lim_{n\to\infty}r_n/n=u>0$. Then
$$
\int_{\zeta}^{-1}(-1-t)^{r_n}t^{s_n}q(t)\,dt=
\frac{ \zeta^{s_n-r_n}q^{(k)}(\zeta) }{ (r_n\zeta-s_n\zeta^{-1})^{k+1}}
+O\left(\frac{1}{n^{k+2}}\right).\tag\secA.11
$$
Furthermore, for any $\epsilon>0$ the implicit constant is uniform for $u\in[\epsilon,\infty)$ and 
independent of $s_n$. \epf

\endproclaim

\medskip
{\it Proof of Theorem \secA.1.} Suppose first that $u<0$. 
Then (2.1) holds for large enough $n$, and for $3|1-q$ \cite{\ec, (7.11)} and \cite{\ec,Lemma 7.4} give
$$
\align
&
\left.\Cal D^l_y\left\{\Cal D^k_x \,P(r_n+x+y,s_n+q(x+y))|_{x=a_1}\right\}\right|_{y=b_1}
\\
&
=
\frac{1}{2\pi i}
\left\langle
  \int_\z^{-1}(-1-t)^{-r_n}t^{-s_n}(t-\z)^{k+l}
  \left\{
    \frac{1}{k!\,l!}(\z-q\z^{-1})^{k+l}(t-\z)^{k+l}
\right.\right.
\\
&\ \ \ \ \ \ \ \ \ \ \ \ \ \ \ \ \ \ \ \ \ \ \ \ \ \ \ \ \ \ \ \ \ \ \ \ \ \ \ \ \ \ \ \ \ \ \ 
\ \ \ \ \ \ \ \ \ \ \ \ \ \ \ \ 
\left.\left.\phantom{\frac{1}{k!\,l!}}
+c_{k+l+1}(t-\z)^{k+l+1}+\cdots
  \right\}
\right\rangle.
\endalign
$$
Proposition \secA.4 applied to the right hand side above yields
$$
\align
&
\left.\Cal D^l_y\left\{\Cal D^k_x \,P(r_n+x+y,s_n+q(x+y))|_{x=a_1}\right\}\right|_{y=b_1}
\\
&\ \ \ \ \ \ \ \ \ \ \ \ \ \ \ \ \ \ \ \ \ \ \ \
\frac{1}{2\pi i}
\left\langle
  \frac{\z^{s_n-r_n}\frac{1}{k!\,l!}(\z-q\z^{-1})^{k+l}(k+l)!}{(r_n\z-s_n\z^{-1})^{k+l+1}}
\right\rangle
+O\left(\frac{1}{n^{k+l+2}}\right),\tag\secA.12
\endalign
$$
which is just what (\secA.1) states. The uniformity of the implicit constant above follows by
the uniformity of the implicit constant in (\secA.11).


Since $(u,v)\neq(0,0)$, at least one of $u<0$, $v<0$, and $-u-v<0$ is true. The 
symmetries $P(\alpha,\beta)=P(-\alpha-\beta-1,\alpha)$ and $P(\alpha,\beta)=P(\beta,\alpha)$
of the coupling function allow one to use the same arguments that proved the case $u<0$ to deduce the 
other two cases (see the proof of Proposition 7.1 in \cite{\ec} for details). \epf

\proclaim{Proposition \secA.5} Let $r_n,s_n\in\Z$ so that $\lim_{n\to\infty}r_n/n=u$ 
and $\lim_{n\to\infty}s_n/n=v$. Then for any integers $k,l\geq0$ and any rational
number $q$ with $3|1-q$ we have
$$
\align
&
\Cal D^k_x \,U_l(r_n+x,s_n+qx)|_{x=a_1}=\frac{1}{2\pi i}{l\choose k}
\left\langle
  \z^{r_n-s_n-1}(1-q\z)^{k}(r_n-s_n\z)^{l-k}
\right\rangle
\\
&\ \ \ \ \ \ \ \ \ \ \ \ \ \ \ \ \ \ \ \ \ \ \ \ \ \ \ \ \ \ \ \ \ \ \ \ \ \ \ \ \ \ \ \ \ \ \ \
\ \ \ \ \ \ \ \ \ \ \ \ \ \ \ \ \ \ \ \ \ \ \ \ \ \
+O\left(n^{l-k-1}\right),\tag\secA.13
\endalign
$$
where $\Cal D^k_x$ acts with respect to some fixed integer sequence $a_1,a_2,\dotsc$. Given any bounded
set $B$ in the plane, the implicit constant can be chosen so that it is uniform 
for $(u,v)\in B$.

\endproclaim

\pf By \cite{\ec,(6.8)} one has
$$
\spreadlines{1\jot}
\align
U_l(a,b)=&\frac{1}{2\pi i}
\left\langle\zeta^{a-b-1}(a-b\zeta)^l\right\rangle\\
&+\ {\text{\rm monomials in $a$ and $b$ of joint degree $<l$}}.
\endalign
$$
This implies, for $3|1-q$, that
$$
\align
&
U_l(r_n+x,s_n+qx)=\frac{1}{2\pi i}
\left\langle
  \z^{r_n-s_n-1}[(1-q\z)x+(r_n-s_n\z)]^l
\right\rangle
\\
&
+\sum_{\alpha,\beta\geq0 \atop \alpha+\beta<l}c_{\alpha,\beta}(r_n+x)^{\alpha}(s_n+qx)^{\beta},
\tag\secA.14
\endalign
$$
where $c_{\alpha,\beta}$ is independent of $r_n$ and $s_n$ for all $\alpha$ and $\beta$.

On the other hand, the argument that proved \cite{\ec,Lemma 6.4} implies that for any constants
$A,B,C\in\C$
$$
\Cal D^k_x(Ax+Br_n+Cs_n)^l|_{x=a_1}={l\choose k}A^k(Br_n+Cs_n)^{l-k}+O\left(n^{l-k-1}\right),
\tag\secA.15
$$
with implicit constant uniform for $(u,v)$ ranging over any bounded set. 
Combining (\secA.14) and (\secA.15) yields the statement of the proposition. \epf

\mysec{5. Interpretation in terms of height functions}

Lozenge tilings of regions with no holes are well known to be interpretable as lattice surfaces 
(see e.g. \cite{\Thurst}\cite{\CLP}). 
Regard the unit triangular lattice $\Cal T$ on which the tiled region lives as being in a horizontal plane, and
let $\Cal L$ be a copy of the lattice $\left(\sqrt{\frac{3}{2}}\Z\right)^3$ placed so that one family of its 
body-diagonals is vertical, intersecting this plane
at the vertices of $\Cal T$. Then each segment joining two nearest neighbors of 
$\Cal L$ projects onto a unit lattice segment of~$\Cal T$.

Orient the lattice segments of $\Cal T$ so that they point in one of the polar directions $\pi/2$, $-\pi/6$,
or $-5\pi/6$.
Then we can lift a lozenge tiling of a lattice region on 
$\Cal T$ by starting from some lattice point, tracing around its tiles one after another, and at each traversal
of a lattice segment $s$ of $\Cal T$, moving either up or down on the corresponding lattice segment of $\Cal L$, 
according as the traversal respected or violated the orientation of~$s$. Tracing around a lozenge results in
going around a lattice square of $\Cal L$ whose orthogonal projection on the plane of $\Cal T$ is that lozenge.

If the tiling has no holes, the node of $\Cal L$ we are finding ourselves at is independent of the
way we traced around the tiles to get there, and what results is a lattice surface in $\Cal L$ whose lattice
square faces are in one to one correspondence with the tiles. 

When a hole of non-zero charge is present this ceases to be true. To illustrate this, suppose we have a
left-pointing triangular hole of side two in our tiling (see Figure 5.1). Then as we trace its boundary 
counterclockwise, the 
traversal of each lattice segment agrees with its orientation. Thus, each complete turn results in six upward
steps along lattice segments of $\Cal L$, and leaves us at the node of $\Cal L$ which is two 
cube-body-diagonals higher than where we started (this is illustrated, from three different viewpoints, 
in Figure 5.2). For a similar right-pointing hole we would end up two such diagonals lower.

In general, if we loop once counterclockwise along a closed walk through the lozenges of a tiling, the 
ending point of the corresponding walk on $\Cal L$ is on the same vertical as its starting point, but
a distance of $q\frac{3}{\sqrt{2}}$ below it, where $q$ is the total charge of the holes we looped around 
(by a negative distance below we mean the absolute value of that distance above). This makes it impossible to 
get a single-sheeted lifting surface with no extra boundary in addition to the the lifting of the boundary
of the region with holes.

Nevertheless, we can construct a {\it multi-sheeted} surface with no additional boundary 
that lifts any
tiling with holes, as follows. Suppose we have a lattice region $\Cal R$ on $\Cal T$ with a finite number of holes, 
and let $T$ be a lozenge tiling of it. For each hole, consider a fixed lattice path 
cut (independent of the tiling $T$) from it to the 
boundary of the region $\Cal R$; 
let $\Cal P$ be the family of lattice paths formed by them.
Whenever a step of a lattice path of $\Cal P$ crosses a lozenge of $T$, 
remove that lozenge from $\Cal R$; let $\Cal R_0$ be the region obtained from $\Cal R$ by removing the union 
$T_0$ of all such lozenges of $T$.
Now regard $T\setminus T_0$ as being a tiling of $\Cal R_0$ in which $\Cal P\cap \Cal R_0$ is part of the boundary. 
Then $T\setminus T_0$ lifts to a single-sheeted lattice 
surface\footnote{\,We assume here that the paths in $\Cal P$ are disjoint.
Unless the holes are packed very close (which, 
since we are interested in scaling limits, will not be the case for us),
$\Cal P$ can be chosen so.
} 
$\bar{S}_T$ in $\Cal L$. The union\footnote{\,For a surface $S$ (single- or multi-sheeted) embedded in $\R^3$
and for $c\in\R$ we denote by $S+c$ the translation of $S$ by the vector $(0,0,c)$.}
$$
U=\bigcup_{n\in\Z}\ \bar{S}_{T}+n\frac{3}{\sqrt{2}}
$$
is a multi-sheeted surface having holes above all the lozenges in $\Cal R\setminus \Cal R_0$. Define $S_T$ to be the
multi-sheeted surface obtained from $U$ by filling in these holes with the missing square faces of the lattice 
$\Cal L$ above the lozenges in $\Cal R\setminus \Cal R_0$.
One readily sees that $S_T$ is independent of the family of cuts $\Cal P$. The surface corresponding to the
tiling in Figure 5.1 is illustrated in Figure 5.3. An instance with three holes is pictured in Figures 5.4
and 5.5.

\topinsert
\centerline{\mypic{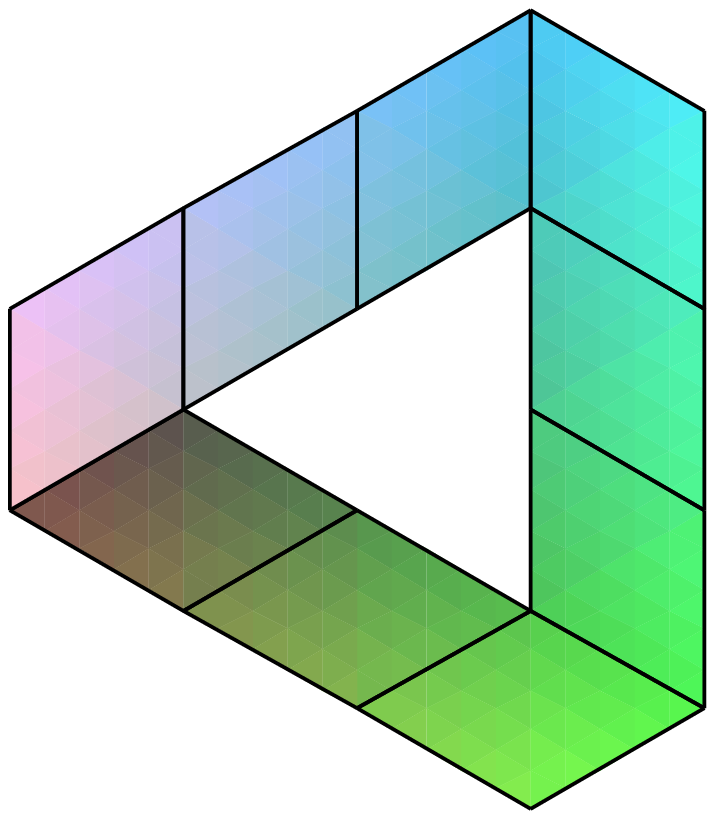}}
\centerline{{\smc Figure~5.1.} {\rm A tiling with a triangular hole of size two.}}
\endinsert

\topinsert
\threeline{\!\!\!\!\!\!\!\!\!\!\!\!\!\!\!\!\!\!\!
\mypic{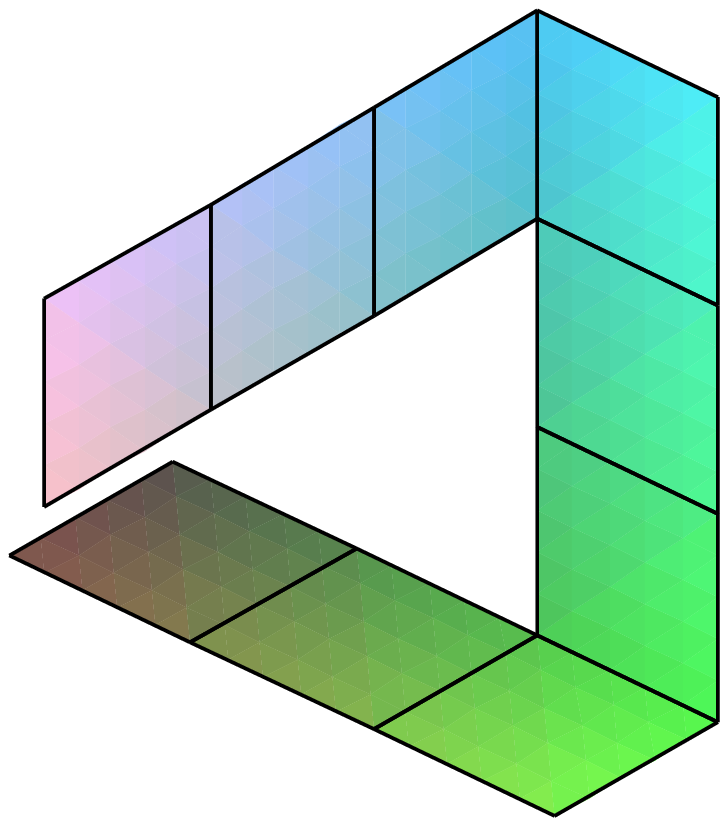}}{\mypic{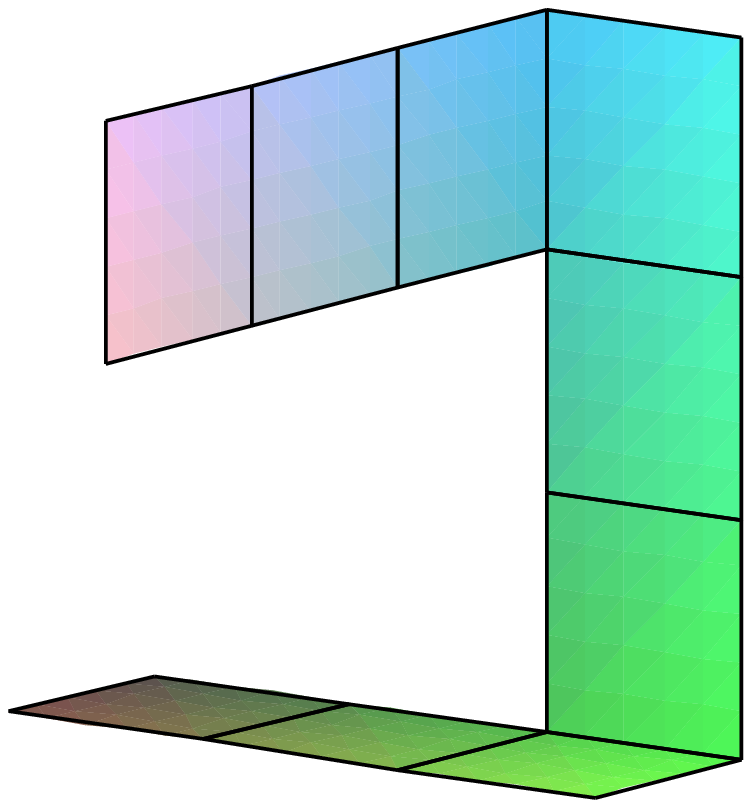}}{\mypic{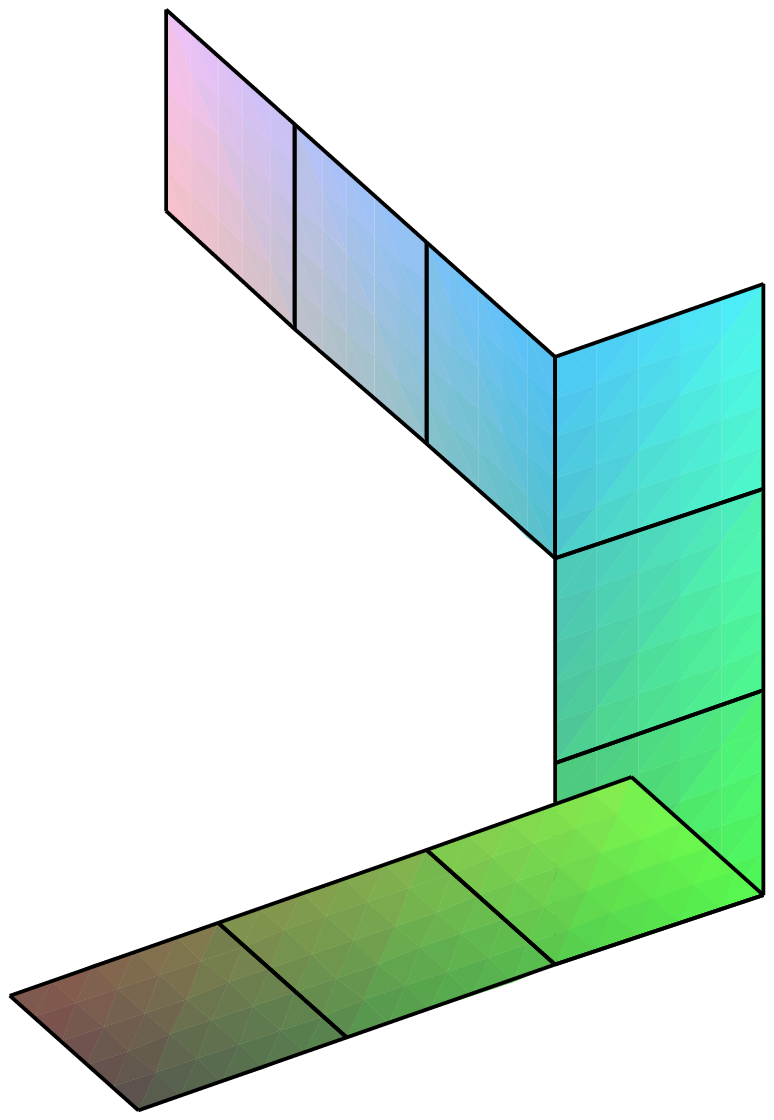}}
\centerline{\smc Figure~5.2. {\rm Three views of $\bar{S}_T$ for the tiling $T$ above.}}
\endinsert


\medskip
The detailed definition of the joint correlation $\hat\omega$ is the following (see \cite{\ec}).
For $j=1,\dotsc,k$, let $Q_j$ be either a lozenge-hole or a lattice triangular 
hole of side two.
 
It is enough to define $\hat\omega(Q_1,\dotsc,Q_k)$ when $q=\sum_{j=1}^k \ch(Q_j)\geq0$ (the other case reduces 
to this by reflection across a vertical lattice line). Our definition is inductive on $q$:

\topinsert
\twoline{\mypic{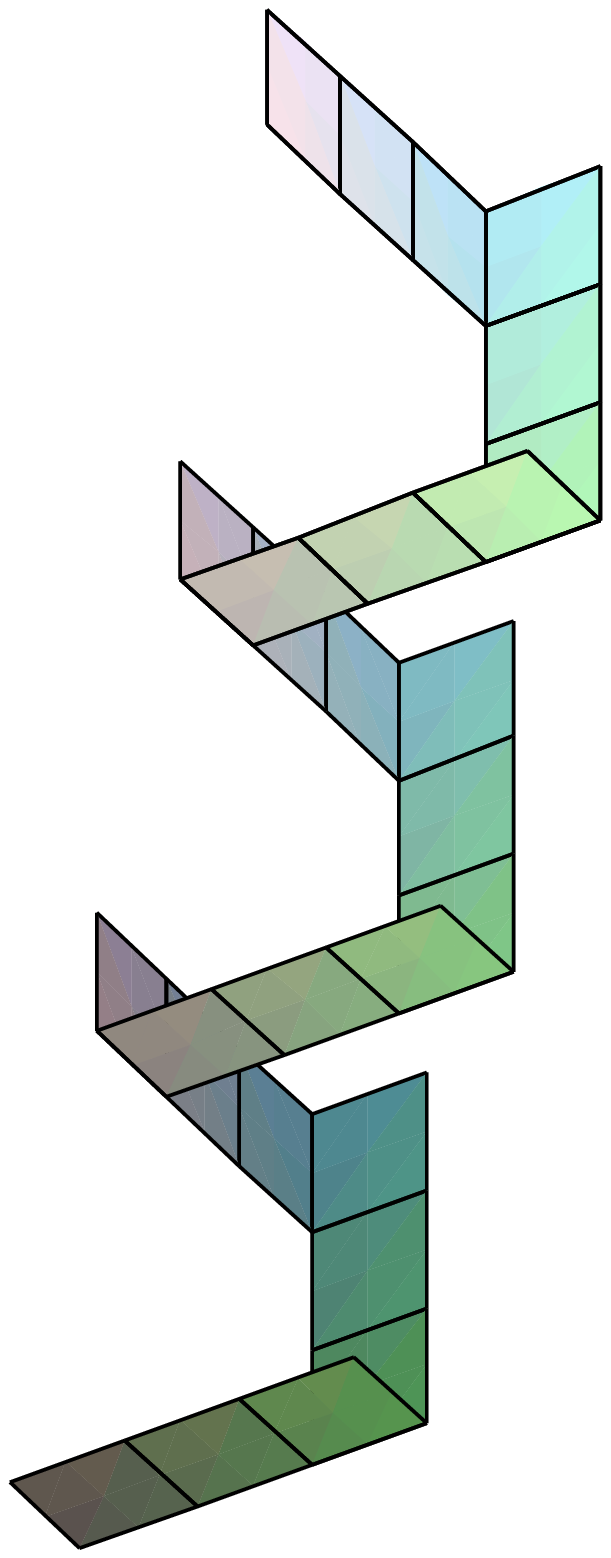}}{\mypic{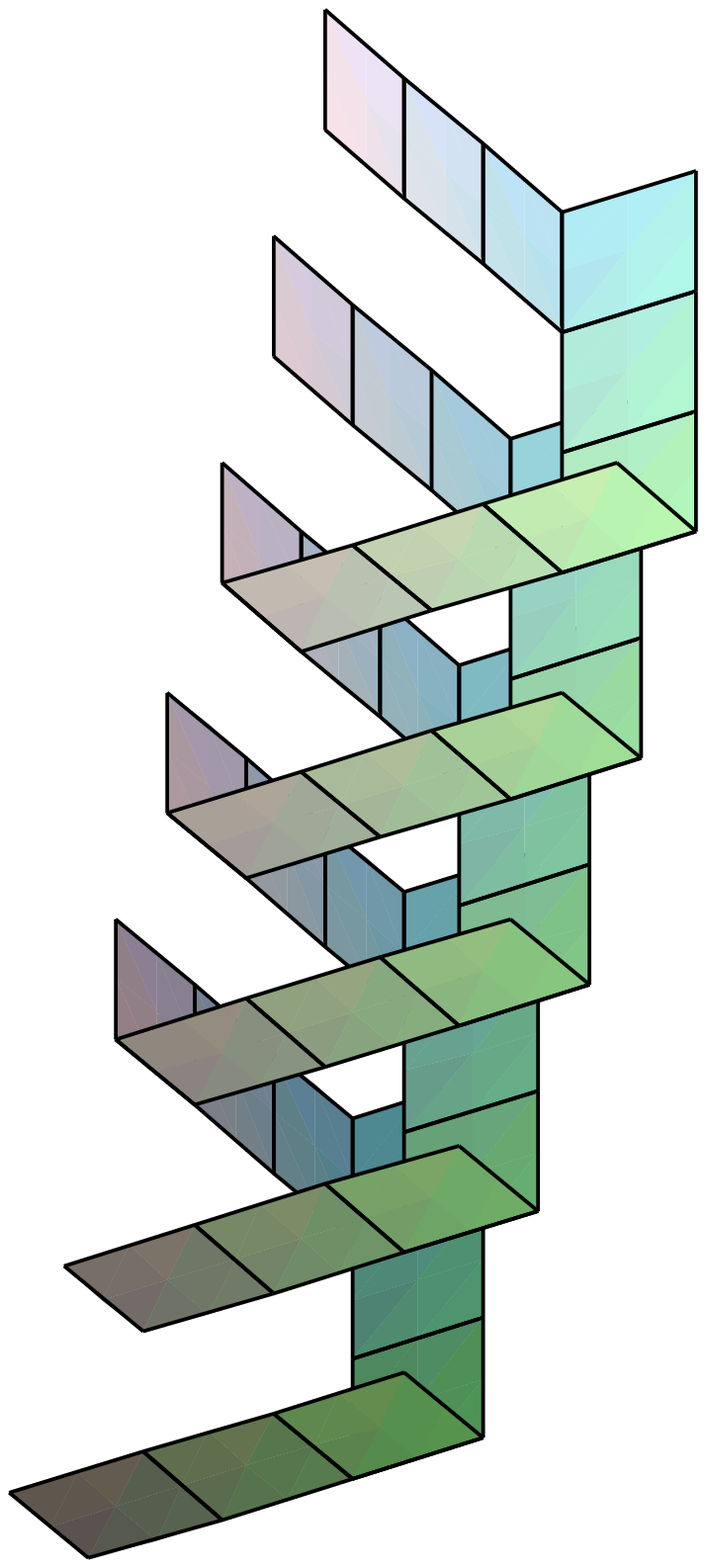}}
\centerline{\smc Figure~5.3. {\rm (a) One of the two connected components of $S_T$ for the tiling in Figure 5.1. 
(b) The full $S_T$. }}
\endinsert

\medskip
$(i)$. If $q=0$, let $N$ be large enough so that the lattice rhombus of side $N$ centered at the origin encloses 
all $Q_j$'s, and denote by $T_N$ the torus obtained from this large lattice rhombus by identifying its opposite 
sides. Set\footnote{\,$\M(\Cal R)$ denotes the number of lozenge tilings of the lattice region $\Cal R$.}
$$
\hat\omega(Q_1,\dotsc,Q_k):=
\lim_{N\to\infty}\frac{\M\left(T_N\setminus Q_1\cup\cdots \cup Q_k\right)}{\M\left(T_N\right)}.\tag5.1
$$

$(ii)$. If $q>0$, define
$$
\hat\omega(Q_1,\dotsc,Q_k):=\lim_{R\to\infty}R^q\,\hat\omega\left(Q_1,\dotsc,Q_k,W_{R,0}\right).\tag5.2
$$

\topinsert
\ltwoline{\!\!\!\!\!\!\!\!\!\!\!\!\!\!\!\!\!\!\!\!\!\!\!\!\!\!\!\!\!\!\!\!\!\!\!\!\!\!\!\!
\mypic{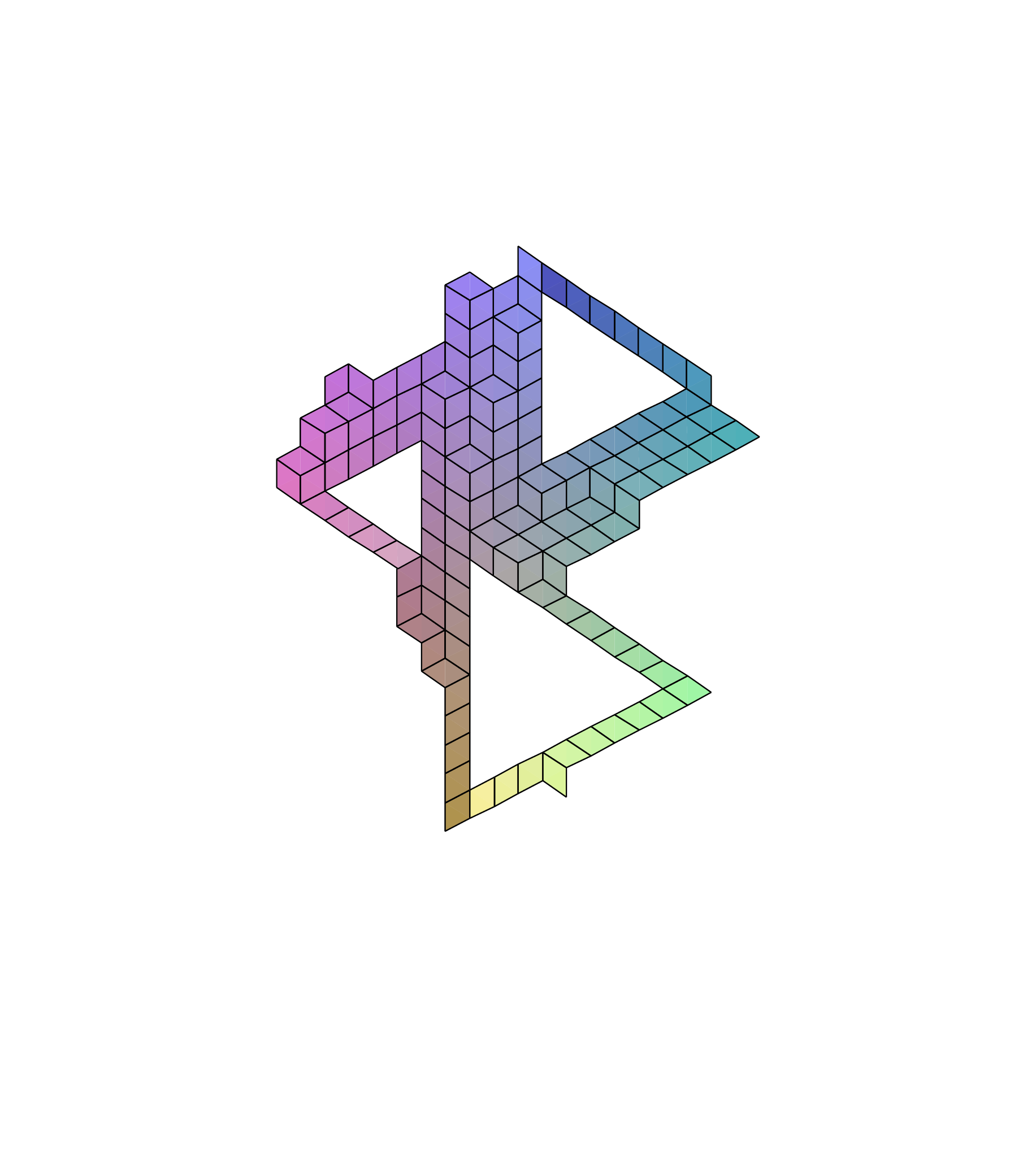}}
{\!\!\!\!\!\!\!\!\!\!\!\!\!\!\!\!\!\!\!\!\!\!\!\!\!\!\!\!\!\!\!\!\!\!\!\!\!\!
\mypic{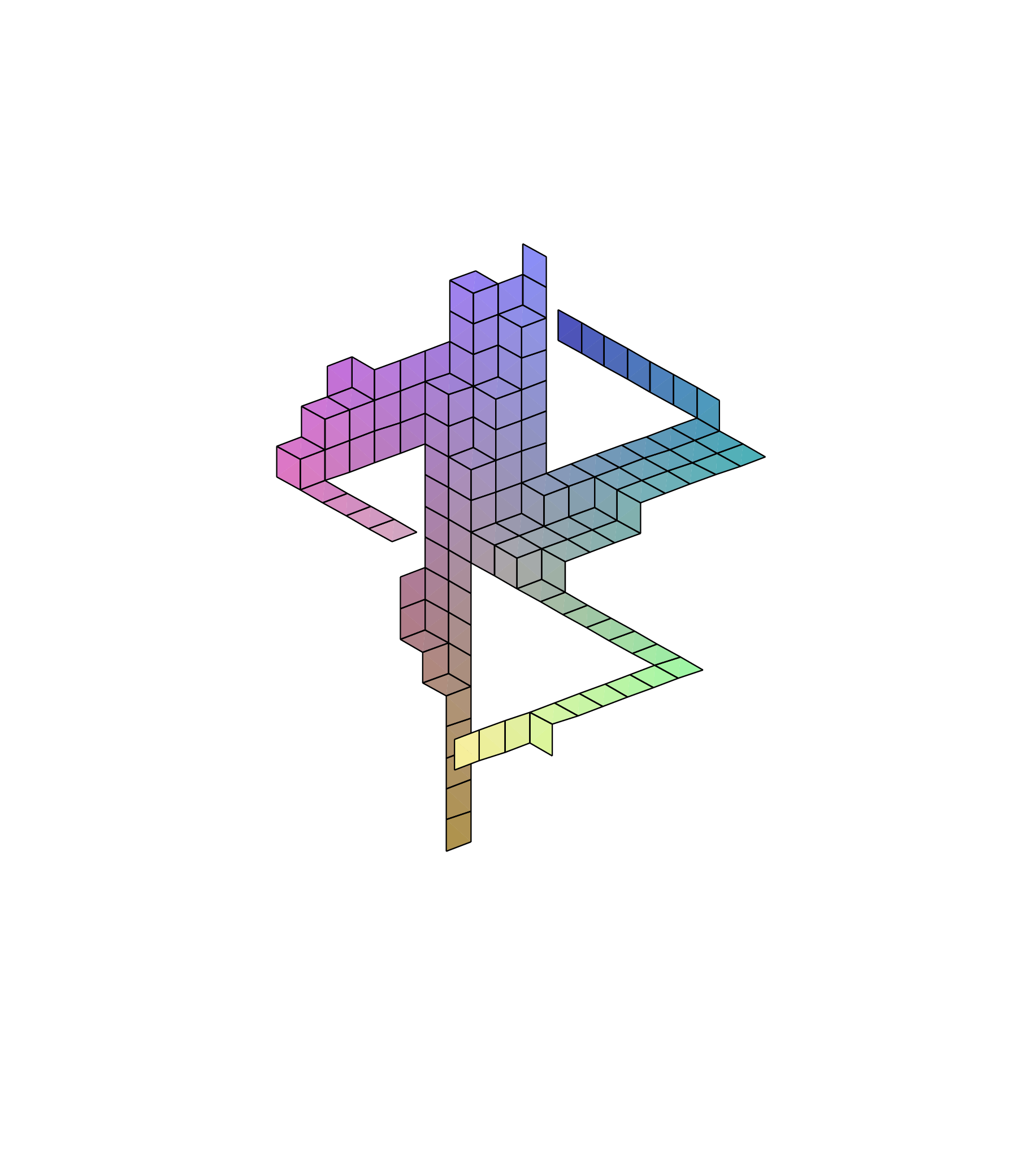}}
\centerline{\smc Figure~5.4. {\rm (a) A tiling $T$ with three holes. (b) A view of $\bar{S}_T$.}}
\endinsert

\topinsert
\ltwoline{\!\!\!\!\!\!\!\!\!\!\!\!\!\!\!\!\!\!\!\!\!\!\!\!\!\!\!\!\!\!\!\!\!\!\!\!\!\!\!\!\!\!\!\!\!\!
\!\!\!\!\!\!
\mypic{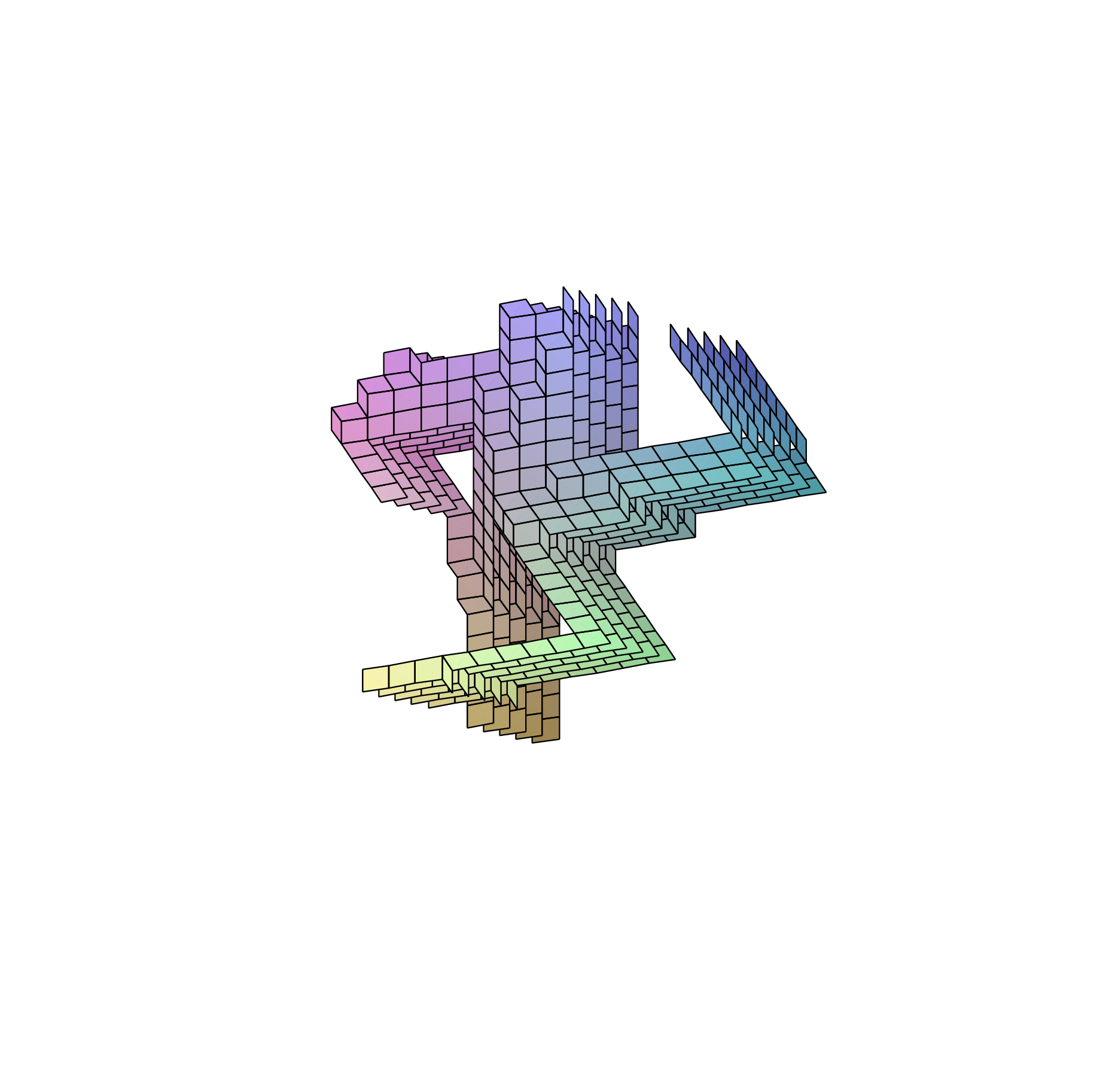}}
{\!\!\!\!\!\!\!\!\!\!\!\!\!\!\!\!\!\!\!\!\!\!\!\!\!\!\!\!\!\!\!\!\!\!\!\!\!\!
\!\!\!\!\!\!\!\!\!\!\!\!\!\!\!

\mypic{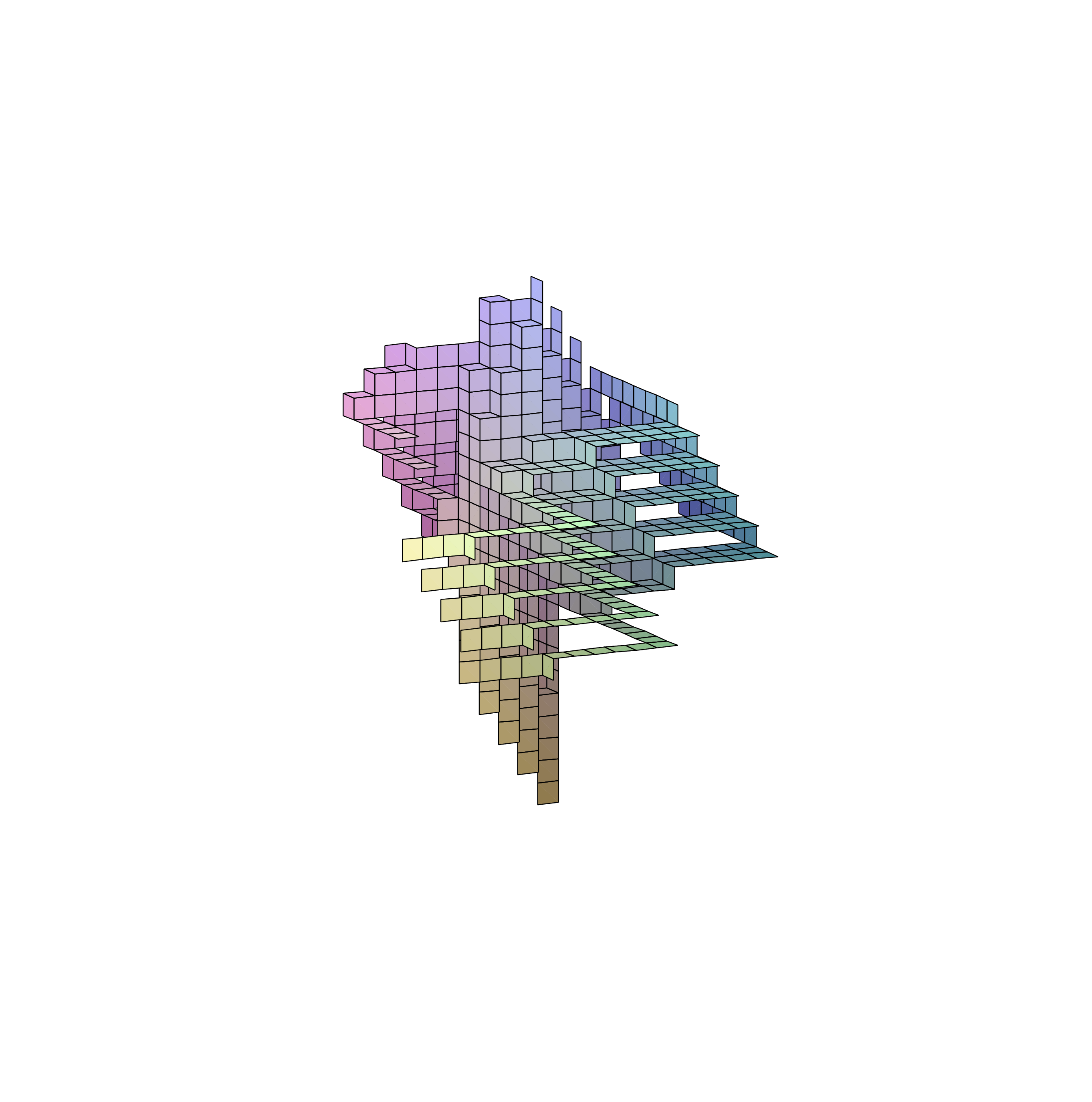}}
\centerline{\smc Figure~5.5. {\rm Two views of $S_T$ for the tiling $T$ in Figure 5.4(a).}}
\endinsert

\medskip
\flushpar
The above limits exist by Proposition 2.1.

Let $\Delta_1,\dotsc,\Delta_k$ be fixed lattice triangular holes of side two, and let $L$ be a fixed 
lozenge position. Assume $q=\sum_{j=1}^k \ch(\Delta_j)=0$.
In the limit measure of the uniform measures on the tori 
$T_N\setminus\Delta_1\cup\dotsc\cup\Delta_k$, the probability that $L$ is occupied by a lozenge in a random 
tiling is
$$
\text{\rm Prob\,\{$L$ is occupied\}}=
\lim_{N\to\infty}
\frac{\M\left(T_N\setminus L\cup\Delta_1\cup\cdots \cup \Delta_k\right)}
{\M\left(T_N\setminus \Delta_1\cup\cdots \cup \Delta_k\right)}
=\frac{\hat\omega(L,\Delta_1,\dotsc,\Delta_k)}{\hat\omega(\Delta_1,\dotsc,\Delta_k)}.\tag5.3
$$
This expression for the probability of $L$ being occupied holds in fact for general $q$. Indeed,
suppose this has been established for total charges $<q$. Then the probability that $L$ is occupied in a random
tiling with the extra hole $W_{R,0}$ in addition to $\Delta_1,\dotsc,\Delta_k$ is
$$
\frac{\hat\omega(L,\Delta_1,\dotsc,\Delta_k,W_{R,0})}{\hat\omega(\Delta_1,\dotsc,\Delta_k,W_{R,0})}.
$$
However, in the limit $R\to\infty$ this is, by (5.2), the same as the fraction on the right hand side of (5.3), 
and our statement is proved by induction.

Suppose $\Cal R$ is a bounded simply connected lattice region on $\Cal T$. Then each tiling of $\Cal R$ lifts
to a single-sheeted lattice surface on $\Cal L$, and we can define the average lifting surface 
$S_{\av}$ by taking the arithmetic mean of the finitely many heights of these surfaces above each node of 
$\Cal T$. As pointed out in \cite{CLP}, if $u$ and $v$ are nearest neighbors 
in $\Cal T$ so that the lattice segment between them is oriented from $u$ to $v$ (see the second paragraph of 
this section), then 
$$
S_{\av}(v)-S_{\av}(u)=\frac{1}{\sqrt{2}}\left(1-3p(L_{uv})\right),\tag5.4
$$
where $L_{uv}$ is the lozenge location whose short diagonal is $uv$ (the factor multiplying the parenthesis on 
the right hand side arises because the traversal of each unit segment in $\Cal T$ results in a change of height 
on the lifting surface of one third of a body diagonal of a lattice cube of $\Cal L$). Thus, the height of 
$S_{\av}$ at any node of $\Cal T$ can be obtained by taking cumulative sums of lozenge occupation probabilities.

The regions we are concerned with --- complements of finite unions of disjoint lattice triangular
holes of side 2
--- have an 
infinite set of lozenge tilings, so we cannot use the arithmetic mean as the definition of the average of the
surfaces their tilings lift to. However, since we know the lozenge placement probabilities are given by (5.3), 
we can turn (5.4) around and use it to define this average surface.

More precisely, consider for each triangular hole a lattice path in $\Cal T$ from a point on its boundary to
infinity. Let $\Cal P$ be the union of these lattice paths; assume they are disjoint. Define $\bar{S}_{\av}$
to be the lattice surface on $\Cal L$ satisfying
$$
\bar{S}_{\av}(v)-\bar{S}_{\av}(u)=\frac{1}{\sqrt{2}}\left(1-3p(L_{uv})\right),\tag5.5
$$
for any two nearest neighbors $u$ and $v$ of $\Cal T$ for which the segment between them is oriented
from $u$ to $v$ and is not crossed by any path of $\Cal P$. Define the average lifting surface of the tilings
of the complement of the holes by
$$
S_{\av}=\bigcup_{n\in\Z}\ \bar{S}_{\av}+n\frac{3}{\sqrt{2}}.
$$
This definition is readily seen to be independent of the family of cuts $\Cal P$.

We note that in the case when there are no holes, the average surface can be defined in terms of the 
translation invariant ergodic measure on the hexagonal lattice (the lattice whose dimer coverings
are equivalent to lozenge tilings of the triangular lattice), which follows by a general result of Sheffield 
(see \cite{\Sheffield}\cite{\KOS})
to be unique and given by a limit of uniform measures on tori. Due to the presence of holes our setting does 
not seem to fit that context.

The helicoid $H(a,b;c)$ is the surface whose parametric equations in Cartesian coordinates are
$$
\align
x&=a+\rho\cos\theta\\
y&=b+\rho\sin\theta,\ \ \ \ \ -\infty<\rho,\theta<\infty.\tag5.6\\
z&=c\theta
\endalign
$$
The half helicoid $H^+(a,b;c)$ is obtained by restricting the range of $\rho$ in the above to $(0,\infty)$.
The dotted helicoid $\dot{H}(a,b;c)$ is $H(a,b;c)$ minus the vertical axis $x=a,y=b$. 

For positive integers $s$ define the $s$-refined half helicoid by
$$
{H}^+_s(a,b;c):=\bigcup_{j=0}^{s-1}\ {H}^+(a,b;c)+\frac{2\pi cj}{s}.\tag5.7
$$
Define the $s$-refined dotted helicoid by
$$
\dot{H}_s(a,b;c):=\bigcup_{j=0}^{s-1}\ \dot{H}(a,b;c)+\frac{\pi cj}{s}.\tag5.8
$$
Note that for $s\in\Z$, the fibers of ${H}^+_{s}(a,b;sc)$ above each point $u$ in the $xy$ coordinate plane 
are of the form $f(u)+2\pi c\Z$. Thus, given ${H}^+_{s_i}(a_i,b_i;s_ic)$, $s_i\in\Z$, $i=1,\dotsc,k$, one can 
define their sum 
$$ 
S:={H}^+_{s_1}(a_1,b_1;s_1c)+\cdots+{H}^+_{s_k}(a_k,b_k;s_kc)
$$
by defining the fiber of $S$ above $u$ to be 
$$
f_1(u)+\cdots+f_k(u)+2\pi c\Z.\tag5.9
$$
We can define the sum of dotted helicoids analogously, using that the fibers of $\dot{H}_s(a,b;sc)$ above $u$
are of the form $f(u)+\pi c\Z$. Figure 5.6(a) illustrates $H^+_2(0,0;2)+H^+(1,0;-1)$. Note that a point on
the surface making a complete counterclockwise turn around the ``spiral stairwell'' on the left ends up two levels
higher; a similar turning around the other spiral takes the point one level lower. Figure 5.6(b) illustrates 
the effect of doubling the refinement indices in a sum of refined half helicoids: the new surface
is the union of the original surface and a suitable vertical translate of it. A depiction of a sum of three
helicoids is given in Figure 5.7.

\topinsert
\ltwoline{\!\!\!\!\!\!\!\!\!\!\!\!\!\!\!\!\!\!\!
\mypic{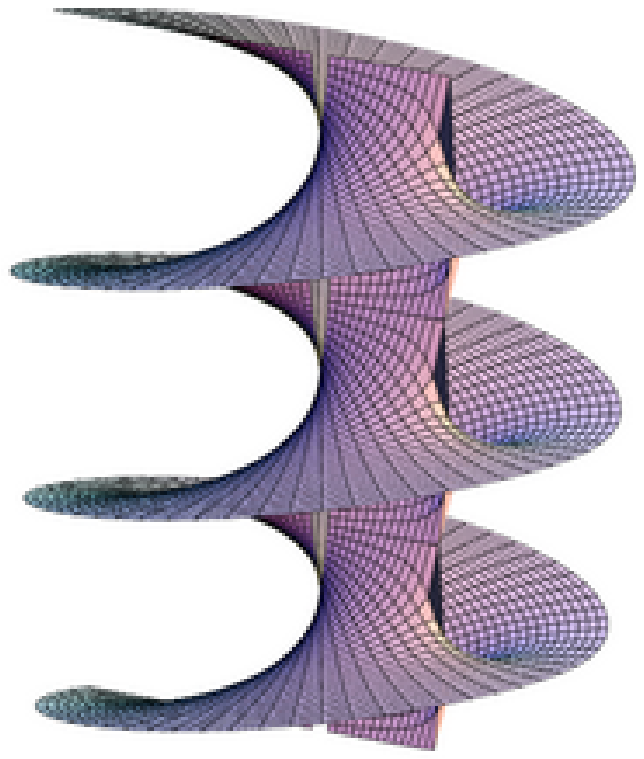}}
{\!\!\!\!\!\!\!\!\!\!\!\!\!\!\!\!\!\!\!\!\!\!\!\!\!\!\!\!\!\!\!\!\!\!\!\!\!\!
\!\!\!\!\!\!\!\!\!\!\!\!\!\!\!\!\!\!\!

\mypic{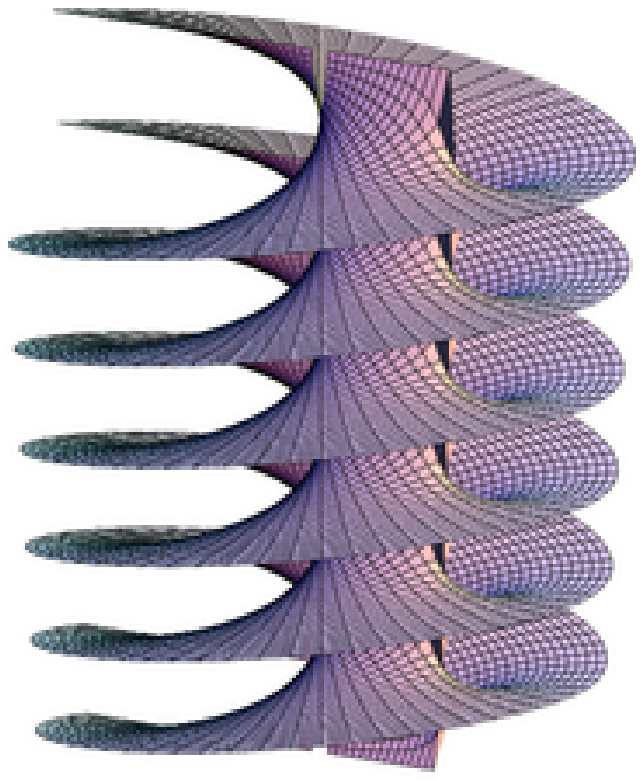}}
\centerline{\smc Figure~5.6. {\rm (a) $H^+_2(0,0;2)+H^+(0,1;-1)$. (b) $H^+_4(0,0;2)+H^+_2(0,1;-1)$ is}}
\centerline{\ \ \ \ \ \ \ \ \ \ obtained by superimposing two copies of the surface (a).}
\endinsert

\topinsert
\ltwoline{\!\!\!\!\!\!\!\!\!\!\!\!\!\!\!\!\!\!\!\!\!\!\!\!\!\!\!\!\!\!\!\!\!\!\!\!\!\!\!\!
\mypic{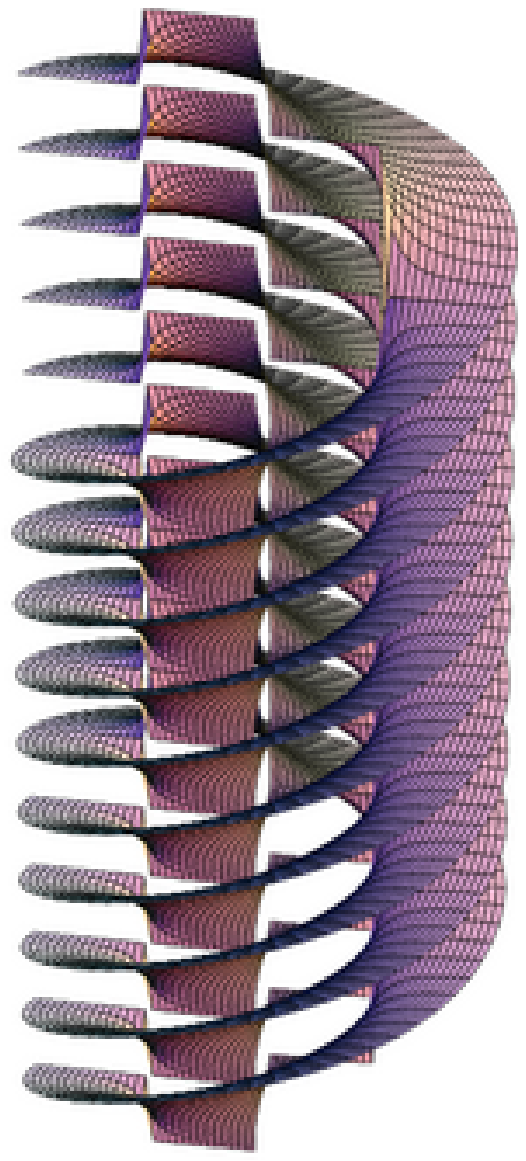}}
{
\mypic{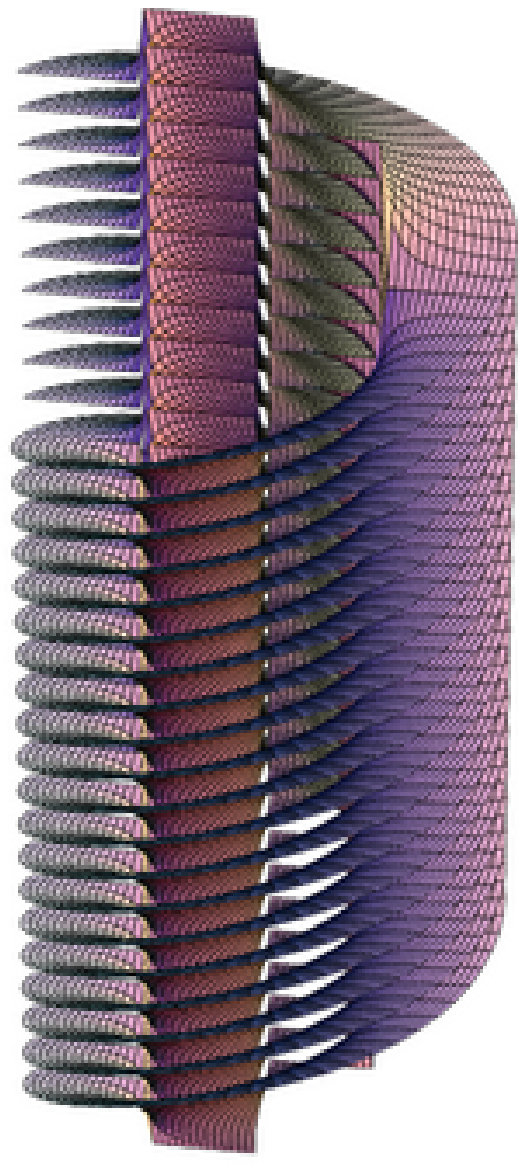}}
\centerline{{\smc Figure~5.7.} 
{\rm (a) The sum of the refined half helicoids $H^+_2(-1,0;-2)$, $H^+_3(0,0;3)$ and $H^+_4(1,0;4)$. (b) Two}}
\centerline{copies of the latter give $H^+_4(-1,0;-2)+H^+_6(0,0;3)+H^+_8(1,0;4)$.}
\endinsert

Let $\Cal T_R$ be the lattice obtained from the triangular lattice $\Cal T$ by a homothety around the 
origin of factor~$\frac1R$.

\proclaim{Theorem 5.1} Let $x_1^{(R)},\dotsc,x_m^{(R)}$, $y_1^{(R)},\dotsc,y_m^{(R)}$,
$z_1^{(R)},\dotsc,z_n^{(R)}$ and $w_1^{(R)},\dotsc,w_n^{(R)}$ be sequences of integers so that
$\lim_{R\to\infty}x_i^{(R)}/R=x_i$, $\lim_{R\to\infty}y_i^{(R)}/R=y_i$, 
$\lim_{R\to\infty}z_j^{(R)}/R=z_j$ and $\lim_{R\to\infty}w_j^{(R)}/R=w_j$ for $1\leq i\leq m$ and
$1\leq j\leq n$. Assume the $(x_i,y_i)$'s and $(z_j,w_j)$'s are all distinct. 

Let ${E}^q_{{\bold a}_1},\dotsc,{E}^q_{{\bold a}_m}$ and 
${W}^q_{{\bold b}_1},\dotsc,{W}^q_{{\bold b}_n}$, $3|1-q$, be multiholes on  $\Cal T_R$, and let the lists
${\bold a}_i$ and ${\bold b}_j$ have lengths $s_i$ and $t_j$, respectively.
Let 
$$
S_{\av}^{\Cal T_R}=S_{\av}^{\Cal T_R}\left({E}^q_{{\bold a}_1}\left(x_1^{(R)},y_1^{(R)}\right),\dotsc,
{W}^q_{{\bold b}_n}\left(z_n^{(R)},w_n^{(R)}\right)\right)
$$
be the average lifting surface of the tilings of the complement of these multiholes on $\Cal T_R$.

Then, as $R\to\infty$, $RS_{\av}^{\Cal T_R}$ converges to the sum of refined helicoids 
$$
\align
&\ \ \ 
\sum_{i=1}^m H^{+}_{2s_i}\left(x_i,y_i;-\frac{3s_i}{\sqrt{2}\pi}\right)
+
\sum_{j=1}^n H^{+}_{2t_j}\left(z_j,w_j;\frac{3t_j}{\sqrt{2}\pi}\right)
\\
&
=
\sum_{i=1}^m \dot{H}_{s_i}\left(x_i,y_i;-\frac{3s_i}{\sqrt{2}\pi}\right)
+
\sum_{j=1}^n \dot{H}_{t_j}\left(z_j,w_j;\frac{3t_j}{\sqrt{2}\pi}\right).\tag5.10
\endalign
$$
For any bounded set $B$ and any open set $U$ containing $(x_1,y_1),\dotsc,(x_m,y_m)$ and 
$(z_1,w_1),\dotsc,(z_n,w_n)$, the convergence is uniform on $B\setminus U$.

\endproclaim

\pf In addition to our $60^\circ$ oblique coordinate system $\Cal O$ in the plane of the unit triangular 
lattice $\Cal T$, 
consider also a Cartesian system of coordinates $\Cal C$ having the origin at the node of $\Cal T$ just below the
origin of $\Cal O$ , the $x$-axis in the polar direction $0$ and
the $y$-axis in the polar direction $\pi/2$. One readily sees that the nodes of $\Cal T$ have $\Cal C$-coordinates
$\left(a\frac{\sqrt{3}}{2},b\frac{1}{2}\right)$, $a,b\in\Z$, $a+b$ even, and that the midpoint of the segment
connecting the nodes $\left(a\frac{\sqrt{3}}{2},b\frac{1}{2}\right)$ and 
$\left(a\frac{\sqrt{3}}{2},(b+2)\frac{1}{2}\right)$ of $\Cal T$ has $\Cal O$-coordinates
$\left(\frac{a-b}{2},\frac{a+b}{2}\right)$.

We will use Cartesian coordinates to specify points at which 
$S_{\av}^{\Cal T_R}$ is evaluated, and oblique coordinates for the lozenges whose placement probabilities come up.

We have by (5.5)
$$
\spreadlines{3\jot}
\align
&
R\bar{S}_{\av}^{\Cal T_R}\left(\frac{x_0^{(R)}}{R}\frac{\sqrt{3}}{2},\frac{y_0^{(R)}+2}{R}\frac{1}{2}\right)
-
R\bar{S}_{\av}^{\Cal T_R}\left(\frac{x_0^{(R)}}{R}\frac{\sqrt{3}}{2},\frac{y_0^{(R)}}{R}\frac{1}{2}\right)
=
\frac{1}{\sqrt{2}}\left(1-3p_1\left(\frac{x_0^{(R)}-y_0^{(R)}}{2R},\frac{x_0^{(R)}+y_0^{(R)}}{2R}\right)\right)
\\
&\ \ \ \ \ \ \ \ \ \ \ \ \ \ \ \ \ \ \ \ \ \ \ \ \ \ \ \ \ \ \ \ \ \ \ \ \ \ \ \ \ \ \ \ \ \ \ \ 
=
\frac{1}{\sqrt{2}}\left(1-3p_1\left(\frac{x_0^{(R)}-y_0^{(R)}}{2},\frac{x_0^{(R)}+y_0^{(R)}}{2}\right)\right),
\tag5.11
\endalign
$$
where $p_1$ is given by (2.12). The second equality holds because placement probabilities are clearly invariant
under scaling. Analogous equations relate the change of $R\bar{S}_{\av}^{\Cal T_R}$ when moving a distance of 
$\frac1R$ from $\left(\frac{x_0^{(R)}}{R}\frac{\sqrt{3}}{2},\frac{y_0^{(R)}}{R}\frac{1}{2}\right)$ 
in the polar directions $\frac{\pi}{2}+k\frac{\pi}{3}$, $k\in\Z$.

The asymptotics of $p_1\left(x_0^{(R)},y_0^{(R)}\right)-p_2\left(x_0^{(R)},y_0^{(R)}\right)$ follows by (3.1) 
and (1.2). One obtains the asymptotics of 
$p_1\left(x_0^{(R)},y_0^{(R)}\right)-p_3\left(x_0^{(R)},y_0^{(R)}\right)$ similarly from (1.3). Adding up the
two and using $p_1+p_2+p_3=1$ yields
$$
\spreadlines{3\jot}
\align
&
1-3p_1\left(x_0^{(R)},y_0^{(R)}\right)=-\frac{3\sqrt{3}}{2\pi}
\left\{\sum_{i=1}^m s_i\frac{x_0-x_i+y_0-y_i}{(x_0-x_i)^2+(x_0-x_i)(y_0-y_i)+(y_0-y_i)^2}
\right.
\\
&\ \ \ \ \ \ \ \ \ \ \ \ \ \ \ \ \ \ \ \ \ \ \ \ \ \ \ \ \ \ \
\left.
-\sum_{j=1}^n t_j\frac{x_0-z_j+y_0-w_j}{(x_0-z_j)^2+(x_0-z_j)(y_0-w_j)+(y_0-w_j)^2}\right\}\frac{1}{R}
+o\left(\frac{1}{R}\right),\tag5.12
\endalign
$$
with the implicit constant uniform for $(x_0,y_0)\in B\setminus U$. Similar expressions follow for the asymptotics
of $p_2$ and $p_3$.

This allows us to deduce that $\lim_{R\to\infty}R\bar{S}_{\av}^{\Cal T_R}$ exists. Indeed, the height of 
$R\bar{S}_{\av}^{\Cal T_R}$ at any node of $\Cal T_R$ can be obtained by taking cumulative sums of (5.11) and 
its two analogs. By (5.12) and its analogs, the resulting error terms have a zero effect in the 
limit $R\to\infty$: even though they do add up, and there are more and more of them as $R$ gets large, since 
the individual errors are small and the factor $\frac1R$ is present, the combined contribution of the error 
terms is small.

Furthermore, we can find the gradient of the limiting surface $\Cal S$ by dividing the left hand side 
of (5.11) by $\frac1R$ --- the distance between the two points where the heights of $R\bar{S}_{\av}^{\Cal T_R}$ 
are examined --- and letting $R\to\infty$. This yields
$$
\frac{\partial\Cal S}{\partial y}\left(x_0\frac{\sqrt{3}}{2},y_0\frac12\right)=
-\frac{3}{\sqrt{2}\pi}
\left\{\sum_{i=1}^ms_i\frac{\frac{\sqrt{3}}{2}(x_0-x_i)}{\frac34(x_0-x_i)^2+\frac14(y_0-y_i)^2}
-\sum_{j=1}^nt_j\frac{\frac{\sqrt{3}}{2}(x_0-z_j)}{\frac34(x_0-z_j)^2+\frac14(y_0-w_j)^2}
\right\},
$$
for $(x_0,y_0)$ different from $(x_1,y_1),\dotsc,(x_m,y_m)$ and $(z_1,w_1),\dotsc,(z_n,w_n)$. After a change
of variables this becomes
$$
\frac{\partial\Cal S}{\partial y}\left(x_0,y_0\right)=
-\frac{3}{\sqrt{2}\pi}
\left\{\sum_{i=1}^ms_i\frac{(x_0-x_i)}{(x_0-x_i)^2+(y_0-y_i)^2}
-\sum_{j=1}^nt_j\frac{(x_0-z_j)}{(x_0-z_j)^2+(y_0-w_j)^2}
\right\}.\tag5.13
$$
The analog of (5.11) corresponding to moving one unit in the polar direction $\frac{\pi}{6}$ yields by a similar
calculation the directional derivative of $\Cal S$ along this polar direction. Writing $\partial/\partial x$ as
the appropriate linear combination of this directional derivative and $\partial/\partial y$ we arrive at
$$
\frac{\partial\Cal S}{\partial x}\left(x_0,y_0\right)=
\frac{3}{\sqrt{2}\pi}
\left\{\sum_{i=1}^ms_i\frac{(y_0-y_i)}{(x_0-x_i)^2+(y_0-y_i)^2}
-\sum_{j=1}^nt_j\frac{(y_0-w_j)}{(x_0-z_j)^2+(y_0-w_j)^2}
\right\}.\tag5.14
$$

On the other hand, the gradient at $(x,y)$ of the half helicoid $H^+(a,b;c)$ --- and hence of any refinement 
of it --- is $\left(-\frac{c(y-b)}{(x-a)^2+(y-b)^2},\frac{c(x-a)}{(x-a)^2+(y-b)^2}\right)$. Thus, if $\Cal H$
is the sum of refined helicoids
$$
\Cal H=\sum_{i=1}^m H^{+}_{2s_i}\left(x_i,y_i;-\frac{3s_i}{\sqrt{2}\pi}\right)
+
\sum_{j=1}^n H^{+}_{2t_j}\left(z_j,w_j;\frac{3t_j}{\sqrt{2}\pi}\right),\tag5.15
$$
we have that the gradients of $\Cal H$ and $\Cal S$ agree on $\R^2\setminus\{(x_1,y_1),\dotsc,(z_n,w_n)\}$.

By construction, each lifting surface of a tiling of the complement of our multiholes has fibers of the form 
$c+\frac{3}{\sqrt{2}}\Z$. Thus so does $R\bar{S}_{\av}^{\Cal T_R}$ for all $R$, and also the limit surface 
$\Cal S$.
The same is true for each summand in (5.15), and therefore for $\Cal H$. Then the surface 
$\Cal S_0=\Cal S -\Cal H$ is well defined by (5.9). 
By construction, as $u$ approaches $(x_1,y_1)$ the fiber of $\Cal S$ above $u$ approaches 
$0+\frac{3}{\sqrt{2}}\Z$. It follows from the definition of $\Cal H$ that the same is true for the fibers 
of $\Cal H$. This, together with the fact that its gradient is zero, 
implies that $\Cal S_0$ is 
the surface $0+\frac{3}{\sqrt{2}}\Z$. So $\Cal S=\Cal H$, justifying the first expression in (5.10). 
The second expression follows using the elementary fact that 
$H^+(a,b;c)\cup \left(H^+(a,b;c)+\frac{c}{2}\right)=\dot{H}(a,b;c)$ 
(an instance of this can be seen in Figure 5.6(a)). The
stated uniformity of the convergence is implied by the uniformity of the error terms in (5.12) and its analogs.
\epf

\mysec{6. Physical interpretation}

The results of \cite{\sc} and \cite{\ec} already show a close connection between random lozenge tilings with
holes and electrostatics: the joint correlation of holes is given, in the scaling limit, by the electrostatic 
energy of the corresponding system of electrical charges. Theorem 1.1 strengthens this connection by showing
that the electrostatic field itself can be viewed as the scaling limit of the discrete field of
the average orientation of lozenges in a random tiling of the complement of the holes.

Following \cite{\sc,\S 2}, suppose one pictures the two dimensional universe as being spatially quantized
consisting of a very fine lattice of triangular quanta of space.

Associate the complement of the holes with the vacuum of empty space. The quantum fluctuations of the vacuum
cause virtual electron-virtual positron pairs to be ceaselessly created and annihilated. Associate 
virtual electrons with left-pointing unit triangles and virtual positrons with right-pointing unit triangles.
Then a lozenge corresponds to the annihilation of a virtual electron-virtual positron pair, and a lozenge tiling
to one way for all virtual electron-virtual positron pairs to annihilate.

Averaging over all possible tilings arises then naturally as corresponding to the Feynman sum over all 
possible ways the virtual electrons and virtual positrons can annihilate in pairs.
Our result states that 
if the distances between the hole-charges and the distances between them and the point where we are measuring the
${\bold F}$-field are much larger than the lattice spacing of the quantas of space, then this field is almost
exactly the same as the electric field. From this perspective, one sees as the result of an exact calculation how
the (two dimensional) electric field emerges from the quantum fluctuations of the vacuum. This view provides an
electric field which very closely approximates the Coulomb field, but is not exactly equal to it: by Theorem 1.1, 
the latter occurs only in the limit when the size of the quantas of space approaches zero. The discrepancy gets
noticeable only at very small distances. This is in agreement with the expectation of some physicists that linear
superposition for the electric field might break down in the subatomic domain 
(see e.g. Jackson \cite{\Jackson, \S I.3}).


In quantum mechanics, when a virtual electron and a virtual positron annihilate, two photons result, which are
sent away in opposite directions. The unit contribution to ${\bold F}(e)$ of each lozenge covering $e$ in a 
tiling could be regarded as corresponding to one of these two photons (this assumes that the common direction of
the two ejected photons is, on average, the straight line connecting the annihilating virtual particles). 
Then Theorem 1.1 states that
in the scaling limit, the thus generated ``jet'' of photons at $e$ averages precisely to the electrostatic field.
This could then be viewed as a ``mechanism'' for the coming about of the electrostatic field.

The discussion in the previous paragraph leaves the second photon unaccounted for. However, underlying in our 
considerations there is a second discrete vector field we have not mentioned yet: the one given by the average
orientation of the lozenge that covers any fixed {\it right}-pointing unit triangle! Denote it by ${\bold F}'$. An
argument similar to the one that proved Theorem 1.1 shows that ${\bold F}'=-{\bold F}$ in the scaling limit.

We conclude by mentioning another, quite different way to define a vector field via random tilings
with holes. Let $Q_1,\dotsc,Q_n$ be a fixed collection of holes on the unit triangular lattice. For any
$x,y,\alpha,\beta\in\Z$ define
$$
T_{\alpha,\beta}(x,y):=\frac{1}{\sqrt{\alpha^2+\alpha\beta+\beta^2}}
\left(\frac{\hat\omega\left(E_{x+\alpha,y+\beta},Q_1,\dotsc,Q_n\right)}{\hat\omega\left(E_{x,y},Q_1,\dotsc,Q_n\right)}-1\right)
$$
(i.e., the hole $E_{x,y}$ plays the role of a ``test charge,'' and the effects of its displacement are recorded).
Then it can be shown (details will appear in a separate paper) that there exists a vector field ${\bold T}$ so 
that in the scaling limit $T_{\alpha,\beta}(x,y)$ is the orthogonal projection of ${\bold T}(x,y)$ onto the 
vector $(\alpha,\beta)$. Furthermore, up to a constant multiple, the field ${\bold T}$ turns out to be the same 
as ${\bold F}$. Thus, unlike in physics where the electric field is defined by means of a test charge, in our
model there are two different ways to define the corresponding field, and one of them does not use test charges.

We note that the agreement in the previous paragraph is not a matter of course, as it does not hold for instance 
on the critical Fisher lattice --- the planar lattice
of equilateral triangles and regular dodecagons where the edges of the triangles have weight 1, and the 
inter-triangle edges have weight $\sqrt{3}$ (this follows from not yet published joint work with David Wilson).

\bigskip
\bigskip

\mysec{References}
{\openup 1\jot \frenchspacing\raggedbottom
\roster

\myref{\ri}
  M. Ciucu, Rotational invariance of quadromer correlations on the hexagonal lattice, {\it Adv. in Math.} {\bf 191} 
(2005), 46-77.

\myref{\sc}
  M. Ciucu, A random tiling model for two dimensional electrostatics, {\it Mem. Amer. Math. Soc.} {\bf 178} (2005),
no. 839, 1--104.

\myref{\ec}
  M. Ciucu, The scaling limit of the correlation of holes on the triangular lattice with periodic boundary 
conditions, {\it Mem. Amer. Math. Soc.}, accepted, to appear (arXiv preprint math-ph/0501071).

\myref{\ov}
  M. Ciucu, Dimer packings with gaps and electrostatics, {\it Proc. Natl. Acad. Sci. USA} 
{\bf 105} (2008), 2766--2772.

\myref{\CEP}
  H. Cohn, N. Elkies, and J. Propp, Local statistics for random domino tilings of the 
Aztec diamond, {\it Duke Math. J.} {\bf 85} (1996), 117-166.

\myref{\CKP}
  H. Cohn, R. Kenyon, and J. Propp, A variational principle for domino tilings, {\it J. Amer. Math. Soc.}  {\bf 14}
(2001), 297--346.

\myref{\CLP}
  H. Cohn, M. Larsen, and J. Propp, The shape of a typical boxed plane partition, 
{\it New York J. of Math.} {\bf 4} (1998), 137--165.

\myref{\ColdingNotices}
  T. H. Colding and W. P. Minicozzi II, Disks that are double spiral staircases, {\it Notices Amer. Math. Soc.}
{\bf  50}  (2003), 327--339.

\myref{\QED} 
  R. P. Feynman, ``QED: The strange theory of light and matter,'' Princeton University Press, Princeton, New Jersey, 
1985.

\myref{\FS} 
  M. E. Fisher and J. Stephenson, Statistical mechanics of dimers on a plane 
lattice. II. Dimer correlations and monomers, {\it Phys. Rev. (2)} {\bf 132} (1963),
1411--1431.

\myref{\Jackson}
  J. D. Jackson, ``Classical Electrodynamics,'' Third Edition, Wiley, New York, 1998.

\myref{\K}
  R. Kenyon, Local statistics of lattice dimers, {\it Ann. Inst. H. Poincar\'e Probab.
  Statist.} {\bf 33} (1997), 591--618. 

\myref{\KOS}
  R. Kenyon, A. Okounkov, and S. Sheffield, Dimers and Amoebae, {\it Ann. of Math.} {\bf 163} (2006), 1019--1056. 

\myref{\Sheffield}
  S. Sheffield, Random Surfaces, {\it Ast\'erisque}, 2005, No. 304. 

\myref{\Thurst} 
  W. P. Thurston, Conway's tiling groups,  {\it Amer. Math. Monthly}  {\bf  97}  (1990), 757--773.

\endroster\par}

\enddocument